\documentclass[a4paper,fleqn,usenatbib]{mnras}

\usepackage[T1]{fontenc}
\usepackage{ae,aecompl}
\usepackage{subfig}

\usepackage{graphicx}	% Including figure files
\usepackage{amsmath}	% Advanced maths commands
\usepackage{amssymb}	% Extra maths symbols
\graphicspath{{}}

\usepackage{xcolor}

\title[Tidally stripped systems]{A universal model for the evolution of tidally stripped systems}

\author[N. E. Drakos et al.]{
	Nicole E. Drakos$^{1} $\thanks{E-mail: ndrakos@ucsc.edu},
	James E. Taylor$^{2,3}$,
	and Andrew J. Benson$^{4}$
	\\
	$^{1}$Department of Astronomy and Astrophysics, University of California, Santa Cruz, 1156 High Street, Santa Cruz, CA 95064, USA \\	
	$^{2}$Department of Physics and Astronomy, University of Waterloo, 200 University Avenue West, Waterloo, ON N2L\,3G1, Canada \\
	$^{3}$Waterloo Centre for Astrophysics, University of Waterloo, 200 University Avenue West, Waterloo, ON N2L\,3G1, Canada \\
	$^{4}$Carnegie Observatories, 813 Santa Barbara Street, Pasadena, CA 91101, USA\\
}

\date{Accepted XXX. Received YYY; in original form ZZZ}

\pubyear{2021}

\begin{document}
	\label{firstpage}
	\pagerange{\pageref{firstpage}--\pageref{lastpage}}
	\maketitle
	
	\begin{abstract}
Accurate models of the structural evolution of dark matter subhaloes, as they orbit within larger systems, are fundamental to understanding the detailed distribution of dark matter at the present day. Numerical simulations of subhalo evolution support the idea that the mass loss associated with tidal stripping is most naturally understood in energy space, with the particles that are the least bound being removed first. Starting from this premise, we recently proposed a zero-parameter ``energy-truncation model" for subhalo evolution. We tested this model with simulations of tidal stripping of satellites with initial NFW profiles, and showed that the energy-truncation model accurately predicts both the mass loss and density profiles. In this work, we apply the model to a variety of Hernquist, Einasto and King profiles. We show that it matches the simulation results quite closely in all cases, indicating that it may serve as a universal model to describe tidally stripped collisionless systems. A key prediction of the energy-truncation model is that the central density of dark matter subhaloes is conserved as they lose mass; this has important implications for dark matter annihilation calculations, and for other observational tests of dark matter.
	\end{abstract}
	
	
	%%%%%%%%%%%%%%%%%%%%%%%%%%%%%%%%%%%%%%%%%%%%%%%%%%
	
	%%%%%%%%%%%%%%%%% BODY OF PAPER %%%%%%%%%%%%%%%%%%

%%%%%%%%%%%%%%%%%%%%%%%%%%%%%%%%%%%%%%%%%%%%%%%%%%
%%%%%%%%%%%%%%%%%%%%%%%%%%%%%%%%%%%%%%%%%%%%%%%%%%
%%%%%%%%%%%%%%%%%%%%%%%%%%%%%%%%%%%%%%%%%%%%%%%%%%	
\section{Introduction} \label{sec:Intro}
%%%%%%%%%%%%%%%%%%%%%%%%%%%%%%%%%%%%%%%%%%%%%%%%%%
%%%%%%%%%%%%%%%%%%%%%%%%%%%%%%%%%%%%%%%%%%%%%%%%%%
%%%%%%%%%%%%%%%%%%%%%%%%%%%%%%%%%%%%%%%%%%%%%%%%%%	

In the $\Lambda$-Cold Dark Matter ($\Lambda$CDM) framework, positive density fluctuations present in the matter distribution at early times gradually break away from the cosmological expansion and collapse into roughly spherical dark matter haloes. These haloes merge hierarchically on progressively larger scales as the Universe expands, giving rise to the galaxy, group and cluster-scale haloes known today.  Halo mergers are relatively ineffective at mixing infalling material, and the central parts of smaller haloes merging into a larger system can survive for many orbits as self-bound subcomponents, or `subhaloes'. 

It is critical to understand precisely how subhaloes evolve after a merger, 
since a number of important tests of dark matter, including dark matter annihilation signals \citep[e.g.][]{stref2019, delos2019b}, gravitational lensing \citep[e.g.][]{limousin2005,baltz2009,sereno2016},
%, galaxy rotation curves  \citep[e.g.][]{}, 
and the imprint of substructure on stellar streams \citep[e.g.][]{carlberg2020, bonaca2020} depend sensitively on the properties of the subhalo population. The exact internal structure and abundance of small subhaloes is highly uncertain, however; in particular, there are concerns that substructure seen in cosmological simulations
may still be significantly affected by artificial (numerical) disruption  \citep[e.g.][]{vandenbosch2018,errani2021}. These uncertainties, together with baryonic effects, contribute to the small-scale  structure problems often cited as one of the main challenges to the $\Lambda$CDM model  \citep[see, e.g. ][for a review of these problems]{bullock2017}.

%isolated simulations
Given the complexity of cosmological structure formation and the relatively poor resolution of large-volume simulations, subhalo evolution has often been studied using simplified simulations of a single (sub)halo evolving in a fixed background potential \citep[e.g.][]{hayashi2003, kazantzidis2004, boylankolchin2007, kampakoglou2007, penarrubia2008a, penarrubia2008b, penarrubia2009,choi2009, penarrubia2010,drakos2017,ogiya2019, delos2019b,errani2021}. These studies have produced a number of models for tidal stripping and structural evolution that describe how mass is lost as a function of radius. These models generally remain empirical, however, i.e. they contain free parameters or functional forms that are adjusted to match specific simulation results. Thus, while accurate for the specific cases considered, these models are restricted to the particular systems simulated \citep[which are typically isotropic systems with NFW profiles; e.g.][]{hayashi2003, green2019}. 

%in energy space
As first pointed out by \cite{choi2009}, tidal stripping of subhaloes is well approximated as a monotonic, outside-in process in energy space. Extending this idea, in \cite{drakos2017}---Paper~I hereafter---we showed that truncation of the particle distribution function (DF) in a dark matter halo, via a `lowering' operation, produces a set of density profiles very similar to those measured for tidally stripped systems in numerical simulations. This approach has several advantages over models based on simulation data alone, as it predicts the evolution of the central regions of haloes below the resolution limit of numerical simulations, and it is easily generalized to other profiles, or to anisotropic systems. It also has a clear physical interpretation, namely that as mass is lost from a subhalo, the energies change more-or-less uniformly within the system, pushing particles close to the threshold for unbinding over this limit, and thus removing them from the system. In subsequent work \cite[][Paper~II hereafter]{drakos2020}, we developed a full model for tidal stripping that combines this lowering approach with a simple estimate of the mass loss rate. The model has no free parameters, and agrees well with simulations of mass loss from systems with NFW profiles. Since then, a number of other authors have successfully studied the evolution of tidally stripped systems using an energy-based approach \citep[e.g][]{stucker2021,amorisco2021,errani2021b}.

%This paper
The aim of the current study is to test whether the energy-truncation approach and the mass-loss model presented in Papers~I and II are valid for a broader range of tidally stripped collisionless systems. The structure of this paper is as follows: first, in Section~\ref{sec:model} we review the energy-based truncation method introduced in Paper~I and the mass-loss model from Paper~II. In Sections~\ref{sec:collis} and Section~\ref{sec:sims} we summarize the collisionless systems we consider in this paper, and the idealized $N$-body simulations we use to test our models, respectively. In Section~\ref{sec:truncprofs} we calculate from energy truncation the expected evolution of the density profiles of these systems, and compare these predictions to the simulations in Section~\ref{sec:univers}. We show our model predictions for the central density of subhaloes in Section~\ref{sec:compare}. Finally, we explore the physical interpretation of this model in Section~\ref{sec:physical}, and discuss our results in Section~\ref{sec:conc}. In a companion paper, we explore the implications of this work for dark matter annihilation and galaxy-galaxy lensing signal predictions  \citep[][in prep]{drakos2022}.

%%%%%%%%%%%%%%%%%%%%%%%%%%%%%%%%%%%%%%%%%%%%%%%%%%
%%%%%%%%%%%%%%%%%%%%%%%%%%%%%%%%%%%%%%%%%%%%%%%%%%
%%%%%%%%%%%%%%%%%%%%%%%%%%%%%%%%%%%%%%%%%%%%%%%%%%		
\section{Review of  energy-truncation model} \label{sec:model}
%%%%%%%%%%%%%%%%%%%%%%%%%%%%%%%%%%%%%%%%%%%%%%%%%%
%%%%%%%%%%%%%%%%%%%%%%%%%%%%%%%%%%%%%%%%%%%%%%%%%%
%%%%%%%%%%%%%%%%%%%%%%%%%%%%%%%%%%%%%%%%%%%%%%%%%%	

In this section we briefly review the energy-based description of tidal truncation introduced in Paper~I, and the mass-loss model developed in Paper~II.

\subsection{Review of distribution functions of isolated spherical systems} \label{sec:ourmodel_I}

Systems of particles can generically be described by a distribution function (DF) $f(\mathbf r,t) = {\rm d} m/{\rm d}r^3{\rm d}v^3$ which specifies the mass per unit volume ${\rm d}r^3{\rm d}v^3$ at a given location in phase space $(\mathbf r,t)$. For isolated, spherically symmetric and isotropic systems, the DF can be written as a function of a single variable $f(\mathbf r,t) = f(r,v) = f(\mathcal{E})$, where  
$\mathcal{E}= \Psi(r) - v^2/2$ is the (conserved) relative energy and $\Psi(r)$ is the relative potential, defined as $\Psi(r) =- \Phi(r) + \Phi_0$. Here $\Phi(r)$ is the usual gravitational potential, while $\Phi_0$ is a reference potential, usually taken to be the value of  $\Phi$ at the outer boundary of the system. Given the sign convention in the definition of the relative energy, it is positive for all bound particles and represents the binding energy needed to remove the particle from the self-bound system. With these definitions, $f(\mathcal{E})>0$ when $\mathcal{E}>0$, and $f(\mathcal{E})=0$ otherwise.

In terms of these quantities, we can calculate the density profile $\rho(r)$ corresponding to a given distribution function
\begin{equation} \label{eq:rho_DFI}
\begin{aligned}
\rho(r) &= 4 \pi \int_0^{\Psi(r)} f(\mathcal{E}) \sqrt{2(\Psi(r)-\mathcal{E})} {\rm d} \mathcal{E} \\
\end{aligned}
\end{equation}
or we can invert this relationship to determine the isotropic distribution function corresponding to a given density profile
\begin{equation} \label{eq:rho_DFII}
\begin{aligned}
f(\mathcal{E}) &= \dfrac{1}{\sqrt{8}\pi^2} \left[ \int_0^\mathcal{E} 
\dfrac{1}{\sqrt{\mathcal{E} - \Psi}} \dfrac{{\rm d}^2 \rho}{{\rm d}\Psi^2} {\rm d} \Psi + \dfrac{1}{\sqrt{\mathcal{E}}} \left( \dfrac{{\rm d}\rho}{{\rm d}\Psi}\right)_{\Psi=0}
	 \right] \,\,\, ,
\end{aligned}
\end{equation}
where $\Phi(r)$ (and thus $\Psi = \Psi(r)$) is calculated from $\rho(r)$, using Poisson's equation $\nabla^2\Phi(r) = 4\pi G\rho(r)$
\citep[see][for the derivation of these results]{binney}.

\subsection{Lowering the DF to represent tidal truncation}\label{sec:ourmodel_II}

%Subhaloes are clearly not isolated throughout their evolution, and the relative energy defined above will vary with time for each individual particle, due to the effects of tidal heating. The approach of Paper~I was to assume that over the course of a full orbit, the relative energies of all particles change by a constant `tidal energy' $\mathcal{E}_T$ that represents the average integrated work done by tidal forces over the orbit. 
Subhaloes are clearly not isolated throughout their evolution, and the relative energy defined above will vary with time for each individual particle, due to changes in the self-bound potential, and the effects of tidal heating. The approach of Paper~I was to assume that over the course of a full orbit, the relative energies of all particles change by a constant `tidal energy' $\mathcal{E}_T$. Considering the system near apocentre, where heating is minimal, it can then be treated as approximately isolated, with $f \sim f(\mathcal{E})$, where the potential used to calculate $\mathcal{E}$ is that of the particles remaining bound at that time, and all the energies have been shifted by $\mathcal{E}_T$ with respect to the previous apocentre.  By that point, tidal stripping will have removed from the system all the particles in a range of $\mathcal{E}$ between zero and  $\mathcal{E}_T$. 

To estimate the resulting adjustment of the distribution function, we `lower' it by a tidal energy $\mathcal{E}_T$, and use this modified form to recover the new, tidally stripped density profile. This method is similar to how the well-known King model for truncated stellar systems was derived by lowering the DF of an isothermal sphere \citep{king1966}. It was originally proposed in \cite{widrow2005} as a way to truncate NFW profiles for use as initial conditions (ICs) in isolated simulations.  

In terms of the original distribution function, the lowered version is defined as
\begin{equation} \label{eq:fE}
f_{\mathcal{E}_{T}}(\mathcal{E}) = 
\begin{cases}
f_0(\mathcal{E}+\mathcal{E}_{\rm T})-f_0(\mathcal{E}_T)& \mathcal{E} \ge 0\\
0& \mathcal{E} \le 0  \,\,\, ,
\end{cases}
\end{equation}
where $f_0(\mathcal{E})$ is the DF of the original system, and $\mathcal{E}_{ T}$ is the truncation or tidal energy.

Given the lowered distribution function, we could in principle calculate the density profile from Equation~\eqref{eq:rho_DFI}, but we would need the relative potential $\Psi(r)$, which itself depends on $\rho(r)$ through Poisson's equation. Substituting the spherical form of Poisson's equation on the left-hand side of Equation~\eqref{eq:rho_DFI}, we can eliminate $\rho$ and write:
\begin{equation} \label{eq:Psi}
\begin{aligned}
\dfrac{ {\rm d}^2{\Psi(r)} }{{\rm d}r^2} + &\dfrac{2}{r} \dfrac{{\rm d} \Psi(r)}{{\rm d}r} \\
&= -16 \pi^2 G \int_0^{\Psi(r)} f_{\mathcal{E}_{T}}(\mathcal{E}) \sqrt{2(\Psi(r)-\mathcal{E})} {\rm d} \mathcal{E}\,. 
\end{aligned}
\end{equation}
This differential equation, together with the boundary conditions $\Psi(0) = \Psi_0(0) - \mathcal{E}_{ T},\ {\rm d}\Psi(0)/{\rm d }r = 0$ (where $\Psi_0(r)$ is the relative potential of the original, un-truncated system) is easily solved by conventional techniques (see Paper~I). Once we have solved for $\Psi(r)$, the truncation radius $r_t$ is given by the condition $\Psi(r_t)=0$, and the truncated density profile $\rho(r)$ can be calculated from Poisson's equation.

\subsection{Predicting the mass-loss rate}\label{sec:ourmodel_III}

As explained in Paper~II, our proposed mass-loss model is based on the Jacobi model for tidal truncation on a circular orbit \citep{binney,taylor2001a}. A tidal or limiting radius $r_{\lim}$ is defined such that:
\begin{equation}\label{eq:rlim}
\begin{aligned}
	\bar{\rho}_{\rm sat} (r_{\lim}) &= \eta_{\rm eff} \bar{\rho}_{\rm H}(R_p) \\
		\end{aligned}
\end{equation}
where $\bar{\rho}_{\rm sat} (r_{\rm lim})$ is the mean density of the satellite interior to the tidal radius and $\bar{\rho}_{\rm H}(R_p)$ is the mean density of the host halo interior to the pericentre of the satellite's orbit. Throughout this paper, we use $R$ when referring to the radius with respect to the centre of the host halo, and $r$ when referring to the radius within the satellite. 

The  tidal radius, $r_{\rm lim}$, and resulting bound mass of the system, $M_{\rm bnd}$, are then given by the following system of equations:
\begin{equation} \label{eq:Mbnd}
\begin{aligned}
M_{\rm bnd} &=\dfrac{4}{3}  r_{\lim}^3 \bar{\rho}_{\rm sat} (r_{\lim}) \\
M_{\rm bnd} &=  \eta_{\rm eff}  M_H (<R_p) \dfrac{r_{\rm lim}^3}{R_p^3} \\
\end{aligned}
\end{equation}
We emphasize that the tidal radius $r_{\rm lim}$ is not the same as the truncation radius $r_t$ defined in the previous section.

The constant, $\eta_{\rm eff}$ is defined as the orbital average of the instantaneous, spherical $\eta$ value defined in \cite{king1962}: 
\begin{equation} \label{eq:etaeff}
	\eta_{\rm eff} =  \dfrac{1}{t_{\rm orb}} \int_0^{t_{\rm orb}} \left(\dfrac{\omega^2}{\omega_c^2} -  \dfrac{1}{\omega_c^2} \dfrac{{\rm d}^2 \Phi_H}{ {\rm d} R^2}  \right) {\rm d}t \,\,\, ,
\end{equation}
where $\omega=|\mathbf{V} \times \mathbf{R}|/R^2$ is the instantaneous angular velocity of the satellite, $\omega_c=GM_H(<R)/R^3$ is the angular velocity of a circular orbit, and $\Phi_H(R)$ is the potential of the host. For a spherically symmetric system, ${\rm d}\Phi_H/{\rm d}R = GM_H(<R)/R^2$
As demonstrated in Paper~II, with this definition of $\eta$, we obtain a good match to the mass-loss rates measured in our simulations without needing to add or adjust any free parameters. Recently, \cite{stucker2021} proposed that a more natural way to truncate tidally stripped profiles in energy space is using the ``boosted potential". We consider $\eta$ determined from the boosted potential in Appendix~\ref{sec:boost}, and compare it to the $\eta$ values found from the energy-truncation model.

Given a tidal radius defined as in Equation~\eqref{eq:rlim}, we assume the mass outside this radius is lost over the course of an orbit. We adjust the tidal energy $\mathcal{E}_{T}$ in Equation~\eqref{eq:fE} until it produces a satellite with this lowered mass. This gives a full model for the stripped system, allowing us to specify the total mass, distribution function and density profile by the end of the orbit. We apply this stripping model at successive apocentres, since this is the point at which the profile is expected to reach equilibrium. As the system continues to orbit, each successive pericentric passage will decrease $r_{\rm lim}$ and increase $\mathcal{E}_{T}$.\footnote{We take advantage of the fact that truncating the profile first at energy $\mathcal{E}_1$, then truncating the new truncated profile a second time at  $\mathcal{E}_2$ is mathematically equivalent to truncating the original profile at  $\mathcal{E}_1+\mathcal{E}_2$. Therefore, though $r_{\rm lim}$ is calculated using the most recent, stripped density profile, the new profile is always calculated by lowering the original, untrucated DF.}

\subsection{Algorithmic description of method}

In practice, given an orbit and a set of initial conditions, the energy-truncation model can be performed using the following steps:

\begin{enumerate}
\item Compute $\eta_{\rm eff}$ by integrating over the orbit using Equation~\eqref{eq:etaeff}.
\item Solve for the bound mass $M_{\rm bnd}$ and tidal radius, $r_{\rm lim}$, using Equation~\eqref{eq:Mbnd}. Note that $r_{\rm lim}$ is not needed in the following steps.
\item Compute the tidal energy that corresponds to $M_{\rm bnd}$. There is a monotonic relationship between $M_{\rm bnd}$, $r_t$ and $\mathcal{E}_T$, and any one of these three variables can be used to uniquely describe the truncated profile. We create a lookup table to map between  $M_{\rm bnd}$ to $\mathcal{E}_T$\footnote{For convenience, Paper I provides an empirical relationship to map between these variables.}.
%Later, in Fig.~\ref{fig:Mbnd_vs_Et.pdf} we demonstrate the relationship between the variables of the models used in this study. 
\item Using $\mathcal{E}_T$, compute the lowered DF in Equation~\eqref{eq:fE}.
\item Calculate $\Psi(r)$ by solving the ODE in Equation~\eqref{eq:Psi}.
\item Compute the corresponding truncated profile $\rho_{\rm sat}$ using Equation~\eqref{eq:rho_DFI}.
\item Go to ii, and repeat for as many orbits as desired.
\end{enumerate}

%%%%%%%%%%%%%%%%%%%%%%%%%%%%%%%%%%%%%%%%%%%%%%%%%%
%%%%%%%%%%%%%%%%%%%%%%%%%%%%%%%%%%%%%%%%%%%%%%%%%%
%%%%%%%%%%%%%%%%%%%%%%%%%%%%%%%%%%%%%%%%%%%%%%%%%%	
\section{Collisionless satellite models} \label{sec:collis}
%%%%%%%%%%%%%%%%%%%%%%%%%%%%%%%%%%%%%%%%%%%%%%%%%%
%%%%%%%%%%%%%%%%%%%%%%%%%%%%%%%%%%%%%%%%%%%%%%%%%%
%%%%%%%%%%%%%%%%%%%%%%%%%%%%%%%%%%%%%%%%%%%%%%%%%%	

The main goal of this paper is to examine how well sharp truncation of the distribution function at a tidal energy, and the mass-loss model reviewed in the previous section, describe the evolution of a range of different collisionless systems\footnote{A collisionless system is one in which interactions between individual particles are negligible, and thus the gravitational force acting on each particle can be treated as a smooth density field rather than a collection of individual particles.}, including common models for CDM haloes such as the NFW, Hernquist, Einasto models, but also models for stellar systems such as the King model. In this section we review the properties of each of these model systems.

\subsection{NFW} \label{sec:NFW}

In Papers~I and II, we restricted our attention to the NFW profile,
\begin{equation}
\rho(r) = \dfrac{\rho_0 r_{\rm s}^3}{r(r+r_{\rm s})^2} \,\,\, ,
\end{equation}
where $\rho_0$ is a characteristic density and $r_{\rm s}$ is the scale radius, describing the point where the logarithmic slope is ${\rm d} \log \rho/ {\rm d} \log r = -2$. 
%In practice, since this total mass of this system diverges with radius, we will consider a truncated version of this profile in our simulations, realized by iteratively removing particles that are unbound outside some finite outer radius, as explained in Papers~I and II.

For convenience, we provide an analytic description of an energy-truncated NFW profile, using the functional form
\begin{equation} \label{eq:NFWfit}
\rho_{\rm NFWT} = \dfrac{e^{-x /y}} {\left[ 1+ (x/y) \right]^a} \rho_{\rm NFW} \, ,
\end{equation}
where $x= r/r_e$ is the radius normalized by an effective radius, $r_e$, and $y = M_{\rm bnd}/M_{\rm NFW}(<r_s)$ is the bound mass of the truncated profile, normalized by the mass of the untruncated profile within the NFW scale radius, $r_s$. This functional form was chosen as we found it empirically yielded a good fit, and conserves the central density as $r \rightarrow 0$.

The parameters $r_e$ and $a$ can be expressed as follows:
\begin{equation}
\begin{aligned}
\log_{10} r_e &= 0.0811 y^3 + 0.358 y^2 + 0.0781 y  -0.201\\
a &= 0.179 y^3 +0.379 y^2  -0.524 y -0.952 \,\,\,.
\end{aligned}
\end{equation}

\begin{figure}
	\includegraphics[width = \columnwidth]{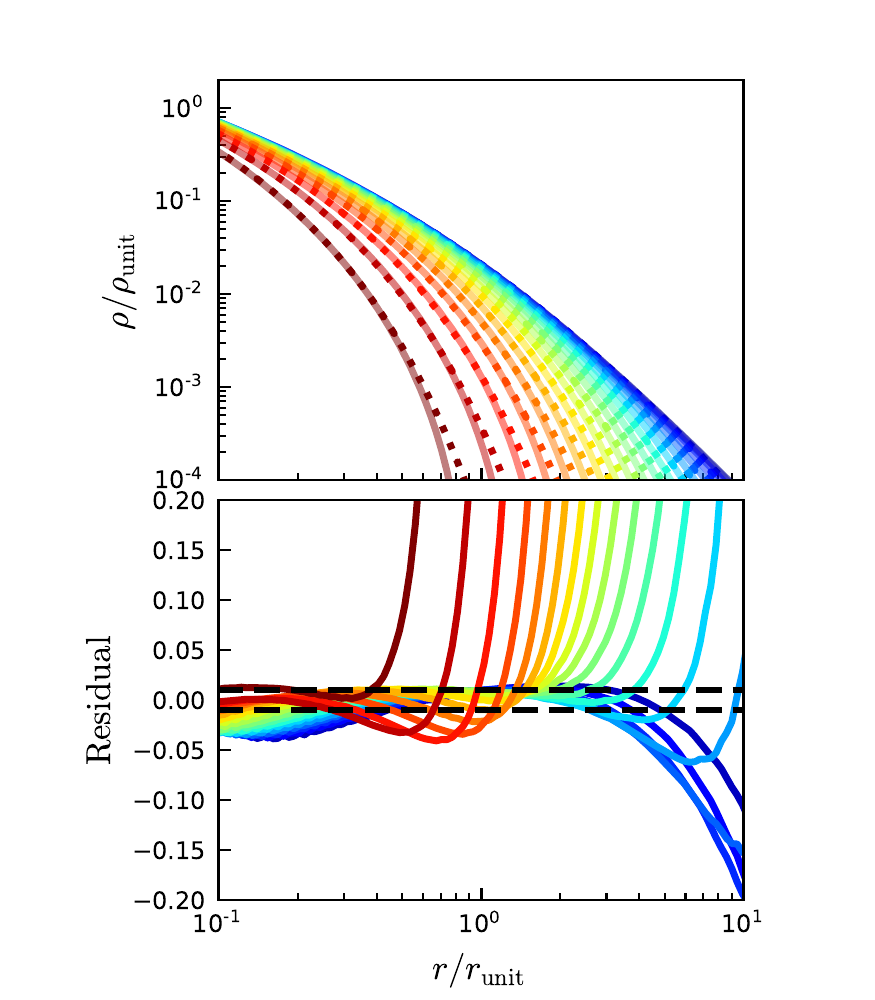}
	\caption{Top: Energy-truncated NFW profiles, $\rho$ (solid lines) and the analytic fit, $\rho_{\rm fit}$ (dotted lines). Bottom: Residuals in the fits, measured as ($\rho_{\rm fit}/\rho - 1)$. The 1 per cent errors are shown with dashed black lines. The analytic fit agrees well with the energy-truncated model inside $\sim 0.5 r_t$.}
	\label{fig:TruncProfs_Fit}
\end{figure}

As shown in Fig.~\ref{fig:TruncProfs_Fit}, this analytic fit generally agrees with the energy-truncated model to within 5 per cent, except at large radii ($r \gtrsim 0.5 r_t$), where the density is very low. Our model predicts a conserved central density, which is captured in Equation~\eqref{eq:NFWfit}; this differs from empirical models calibrated to simulations, which typically include an explicit drop in the central density \citep[e.g.][]{hayashi2003,green2019}. This point will be discussed further in Section~\ref{sec:compare}.

\subsection{Hernquist}

 The Hernquist profile \citep{hernquist1990} was originally used to describe spherical galaxies, but it is also a reasonable approximation for cosmological dark matter halo profiles. The density profile is given by:
 \begin{equation}
 \rho(r)=\dfrac{M_{\rm tot}}{2\pi} \dfrac{a}{r(r+a)^3} \,\,\, ,
 \end{equation}
 where $M_{\rm tot}$ is the total mass and $a$ is a characteristic radius, that encloses a mass of $M_{\rm tot}/4$. 
 
 The advantage to this model is that it has simple analytic expressions for many of its properties, including its DF:
\begin{subequations} 
 \begin{align}
 f(\mathcal{E}) &= \dfrac{M}{8 \sqrt{2} \pi^3 a^3 v_g^3}\dfrac{1}{(1-q^2)^{5/2}}\\
\notag &\hspace{0.5cm} \times \left(3 \sin^{-1}q + q(1-q^2)^{1/2} (1-2q^2)(8q^4-8q^2-3) \right)\\
 q&=\sqrt{\dfrac{a\mathcal{E}}{GM}}\\
 v_g &= \left( \dfrac{GM}{a}\right)^{1/2} \,\,\, .
 \end{align}
 \end{subequations}

\subsection{Einasto}

The Einasto profile was first used to describe star counts in the Milky Way \citep{Einasto1965}. However, this profile is often a better description of cosmological dark matter halo profiles than the well-known NFW profile \citep[e.g.][]{navarro2004, gao2008,klypin2016}. The Einasto profile has the following form:
\begin{equation}
	\begin{aligned}
		\rho(r) 
		&= \rho_{-2} \exp \left( -\dfrac{2}{\alpha}\left[ \left(\dfrac{r}{r_{-2}}\right)^\alpha - 1 \right]\right) \,\,\,,
	\end{aligned}
\end{equation}
where $\alpha$ is the Einasto shape parameter and $r_{-2}$ is the radius where the logarithmic slope is $-2$. Compared to the NFW profile, the Einasto profile has an extra parameter, $\alpha$, that controls the inner slope of the density profile, and may reflect the mass accretion history of the halo \citep[e.g.][]{klypin2016}.

\subsection{King}

The King model resembles an isothermal sphere at small radii, but has a finite mass within a well-defined tidal radius. This model is typically used to describe truncated stellar systems such as globular clusters or elliptical galaxies \citep{king1966}. The King model is derived by lowering the DF of an isothermal sphere:
\begin{equation}
f(\mathcal{E}) = \rho_1 (2 \pi \sigma^2)^{-3/2} (e^{\mathcal{E}/\sigma^2}-1) \,\,\, ,
\end{equation}
where $\sigma$ is the velocity dispersion, $\rho_1$ is a characteristic density.

The density profile of the King model can then be calculated from the DF using Equation~\eqref{eq:rho_DFI}, which gives:
\begin{equation}
\rho(\Psi)=\rho_1\left[ e^{\Psi / \sigma^2} \mbox{erf} \left( \dfrac{\sqrt{\Psi}}{\sigma} \right) - \sqrt{\dfrac{4\Psi}{\pi \sigma^2}}\left( 1 +\dfrac{2 \Psi}{3 \sigma^2} \right) \right] \,\,\, .
\end{equation}

To relate the relative potential energy, $\Psi(r)$, to the density of the profile, $\Psi(r)$ can be solved numerically using Poisson's equation (Equation~\eqref{eq:Psi}). There are many possible parameterizations of the King model, but it can be uniquely defined by the total mass, tidal radius, $r_{\rm t}$ and a dimensionless central potential $P_0= \Psi(0)/\sigma^2$. Alternately, King models can also be characterized by a total mass, tidal radius and a concentration parameter, $c_K$, which depends on the `King radius' $r_0$. The latter quantities are defined as:
\begin{equation}
\begin{aligned}
c_K &= \log_{10} \left( \dfrac{r_{\rm t}}{r_0}\right) \\
r_0 &= \sqrt{\dfrac{9\sigma^2}{4 \pi G \rho_0}} \,\,\, .
\end{aligned}
\end{equation}
where $\rho_0$ is the central density of the halo.

%%%%%%%%%%%%%%%%%%%%%%%%%%%%%%%%%%%%%%%%%%%%%%%%%%
%%%%%%%%%%%%%%%%%%%%%%%%%%%%%%%%%%%%%%%%%%%%%%%%%%
%%%%%%%%%%%%%%%%%%%%%%%%%%%%%%%%%%%%%%%%%%%%%%%%%%	
\section{Simulations} \label{sec:sims}
%%%%%%%%%%%%%%%%%%%%%%%%%%%%%%%%%%%%%%%%%%%%%%%%%%
%%%%%%%%%%%%%%%%%%%%%%%%%%%%%%%%%%%%%%%%%%%%%%%%%%
%%%%%%%%%%%%%%%%%%%%%%%%%%%%%%%%%%%%%%%%%%%%%%%%%%	

The simulations were performed using a version of the $N$-body code \textsc{gadget-2} \citep{gadget2}, modified to contain a fixed background potential corresponding to the host halo.

\subsection{Initial halo models}

We considered four different satellite models for our initial conditions (ICs), as described in the previous section. The ICs were created using our public code \textsc{Icicle} \citep{drakos2017}.  For all models, we define the mass unit to be the initial mass of the satellite, $M_{\rm unit} = M_{\rm sat}$. We define the length unit to be $r_{\rm unit} = a$ for the Hernquist model, $r_{\rm unit} = r_{-2}$ for the Einasto models, and $r_{\rm unit} = 0.1\,r_{\rm t}$ for the King models. Then the density, time and energy units are $\rho_{\rm unit} = M_{\rm unit} r_{\rm unit}^{-3}$, $t_{\rm unit} = \sqrt{r_{\rm unit}^3/G M_{\rm unit}}$ and $E_{\rm unit}=GM_{\rm unit}/r_{\rm unit}$, respectively. We calculated the softening length for each profile as $\epsilon=0.5 r_h N^{-1/3}$,where $r_h$ is the radius enclosing half of the total mass, as in \cite{van2000}.

\begin{figure}
	\includegraphics[width = \columnwidth]{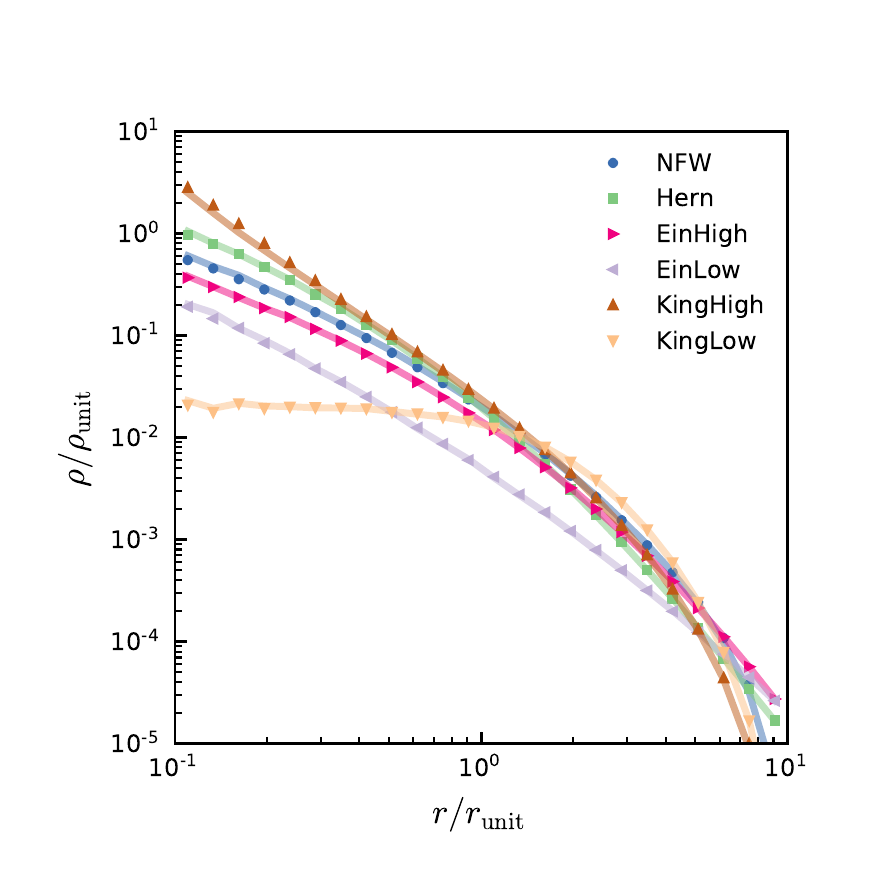}
	\caption{Stability of ICs. Solid lines are the profile at $t=0\,t_{\rm unit}$; points are the profile at  $t=1000\,t_{\rm unit}$.}
	\label{fig:IC_Stability}
\end{figure}

For the NFW profile, we used the same set of simulations described in Paper~I. Since NFW profiles are infinitely extended with divergent mass as $r$ goes to infinity, these ICs were truncated as described in Paper~I. The resulting models resemble a truncated NFW profile with an energy truncation of $\mathcal{E}_{\rm T}\approx0.27$. Though the simulation results can be scaled by the length and mass units, for the Einasto models, there is one free parameter, $\alpha$, to fix; we chose values of $\alpha=0.15$ and $0.3$, as these are representative of the range found in simulations \citep{gao2008}. The King models also have an additional free parameter (which could be specified as $r_0$, $P_0$ or $c$); we used $P_0=3$ and $P_0=12$, which correspond to King concentrations of $c_K=0.7$ and $c_K=2.7$. These concentrations are typical of globular clusters and elliptical galaxies, respectively \citep{binney}. A summary of the IC properties are given in Table~\ref{tab:ICs}. To check the stability of the ICs, they were evolved for $t=1000\,t_{\rm unit}$ in isolation using the N-body code \textsc{gadget-2} \citep{gadget2}, as shown in Fig.~\ref{fig:IC_Stability}. All four profiles are extremely stable outside $r_{\rm unit}=0.1$ at $t=1000\,t_{\rm unit}$.

\begin{table*} 
	\caption{\label{tab:ICs}Summary of the parameters used for the ICs; all satellites have an initial mass of $M_{\rm sat}=1$. The columns list (1) the name of the ICs, (2) the number of particles, (3) the radial unit, (4) specified profile parameters, (5) derived profile parameters,  and (6) \textsc{Gadget-2} softening length,  roughly equivalent to the Plummer softening length. } 
	\begin{tabular}{c c c c c c}
		\hline
		Profile Name & $N$&$r_{\rm unit}$ & IC Parameters & Derived Parameters &$\epsilon/r_{\rm unit}$\\ 
		\hline
		NFWT & $\approx 1.3 \times 10^6$&$r_s$& $r_{\rm cut} = 10\, r_{\rm unit}$ &$\rho_0=0.08\, \rho_{\rm unit}$ & $0.01$ \\
		Hern & $10^6$&$a$& ----- &-----  & $0.01$\\
		EinHigh & $10^6$&$r_{-2}$& $\alpha_E=0.3$& $\rho_{-2}=0.01\, \rho_{\rm unit}$ & $0.02$\\
		EinLow& $10^6$ &$r_{-2}$& $\alpha_E=0.15$ & $\rho_{-2}=0.005\, \rho_{\rm unit}$ & $0.07$\\
		KingHigh& $10^6$& $r_{\rm t}/10$& $P_0=12$ & $\rho_1=0.003\, \rho_{\rm unit}$, $r_0=0.02\, r_{\rm unit}$, $c_K=0.7$&  $0.01$\\	
		KingLow& $10^6$& $r_{\rm t}/10$ & $P_0=3$ &$\rho_1=0.001\, \rho_{\rm unit}$, $r_0=2.13\, r_{\rm unit}$, $c_K=2.7$ &  $0.01$\\		
		\hline
	\end{tabular}	
\end{table*}

\subsection{Satellite orbits} \label{sec:orbits}

 Each satellite model was evolved on two different orbits, corresponding to the `Fast' and `Slow' Simulations in Paper~II (Simulations 3 and 4 from Paper~I), which are representative of orbits that lose mass quickly and slowly, respectively. In Papers I and II we found that the sharp energy truncation model is a better descriptor of orbits which are losing mass slowly. The infinitely extended host halo was assumed to have an NFW profile for all simulations, with a mass $M_{\rm host}=300 M_{\rm unit}$ and a scale radius of $6.69\,r_{\rm unit}$. The orbits in the Slow and Fast Simulations have an apocentre of $r_a = 300\,r_{\rm unit}$, and pericentres of  $R_p=50\,r_{\rm unit}$ and $10\,r_{\rm unit}$, respectively.  The simulations are summarized in Table~\ref{tab:sims}.

\begin{table*}
	\caption{\label{tab:sims} Summary of orbital parameters for the Fast and Slow Simulations. Columns give (1) the simulation name  (2) the apocentric distance (3) the pericentric distance (4) the tangential velocity at apocentre (5) the tangential velocity at pericentre (6) the (radial) orbital period (7) mass-loss factor, $\eta_{\rm eff}$}.
	\begin{tabular}{ c c c c c c c}
		\hline
		 Simulation Name& $R_a/r_{\rm unit}$ & $R_p/r_{\rm unit}$ &$V_a/V_{\rm unit}$ & $V_p/v_{\rm unit}$ & $t_{\rm orb}/t_{\rm unit}$ &  $\eta_{\rm eff}$\\ \hline
		Slow Simulation & 100 & 50 & 1.42 & 2.84 & 185.4 & 2.34 \\
		Fast  Simulation & 100 & 10 & 0.51 & 5.10 &129.7   & 1.72\\
		\hline
	\end{tabular}	
\end{table*}

Subhalo centres and bound particles were identified similarly to the method described in Papers~I and II. Considering only particles that were bound in the previous snapshot, the centre of the satellite was defined as the densest point in position and velocity space. The densest point was calculated using the centre of mass (COM) in progressively smaller spheres, as originally described in \cite{tormen1997}.  Initially, the sphere was centred on the COM of all particles, and the radius was defined as the distance to the furthest particles. The radius was decreased by 90 per cent and recentered on the COM, until there were fewer than 100 particles in the sphere. The velocity frame was calculated in the same way, by using progressively smaller spheres to find the densest point in velocity space. 

Once the frame of the satellite was defined, we found the self-bound particles. First, the energy of each particle was calculated in this frame, assuming a spherical potential (in Paper~II we showed the results are insensitive to this assumption):
\begin{equation}
P_i \approx -Gm \left( \dfrac{N(<r_i)}{r_i} + \sum_{j,r_j>r_i}^N \dfrac{1}{r_j} \right) \,\,\, ,
\end{equation}
where $r_i$ is the distance of particle $i$ from the center of the system. Particles were iteratively removed, and energies recalculated, until convergence. The bound satellite mass is the mass of the remaining bound particles.

%%%%%%%%%%%%%%%%%%%%%%%%%%%%%%%%%%%%%%%%%%%%%%%%%%
%%%%%%%%%%%%%%%%%%%%%%%%%%%%%%%%%%%%%%%%%%%%%%%%%%
%%%%%%%%%%%%%%%%%%%%%%%%%%%%%%%%%%%%%%%%%%%%%%%%%%	
\section{Predicted profile evolution} \label{sec:truncprofs}
%%%%%%%%%%%%%%%%%%%%%%%%%%%%%%%%%%%%%%%%%%%%%%%%%%
%%%%%%%%%%%%%%%%%%%%%%%%%%%%%%%%%%%%%%%%%%%%%%%%%%
%%%%%%%%%%%%%%%%%%%%%%%%%%%%%%%%%%%%%%%%%%%%%%%%%%	

In the energy truncation approach described in Section~\ref{sec:ourmodel_I},  the evolution of the stripped system depends on a single parameter, which can be expressed as the tidal energy $\mathcal{E}_T$. First, we explore how the NFW, Hernquist, Einasto and King models evolve when truncated as function of $\mathcal{E}_T$. We note that applying energy truncation to a King model simply results in another King model, with parameters modified as follows:
\begin{equation}
\begin{aligned}
P_{0,T} &= P_0 - E_T \\
\rho_{1,T} &= \rho_1 \exp(E_T) \\
r_{0,T} & = r_0 \sqrt{\dfrac{\rho(P_0)}{\rho_T(P_{0,T})}} \,\,\, .
\end{aligned}
\end{equation}
For the other profiles, the truncated version no longer has a simple analytic description, but can be calculated from the lowered distribution function. For convenience, we provided an analytic fit to our lowered NFW profile in Section~\ref{sec:NFW}.

\begin{figure}
	\includegraphics[width = \columnwidth]{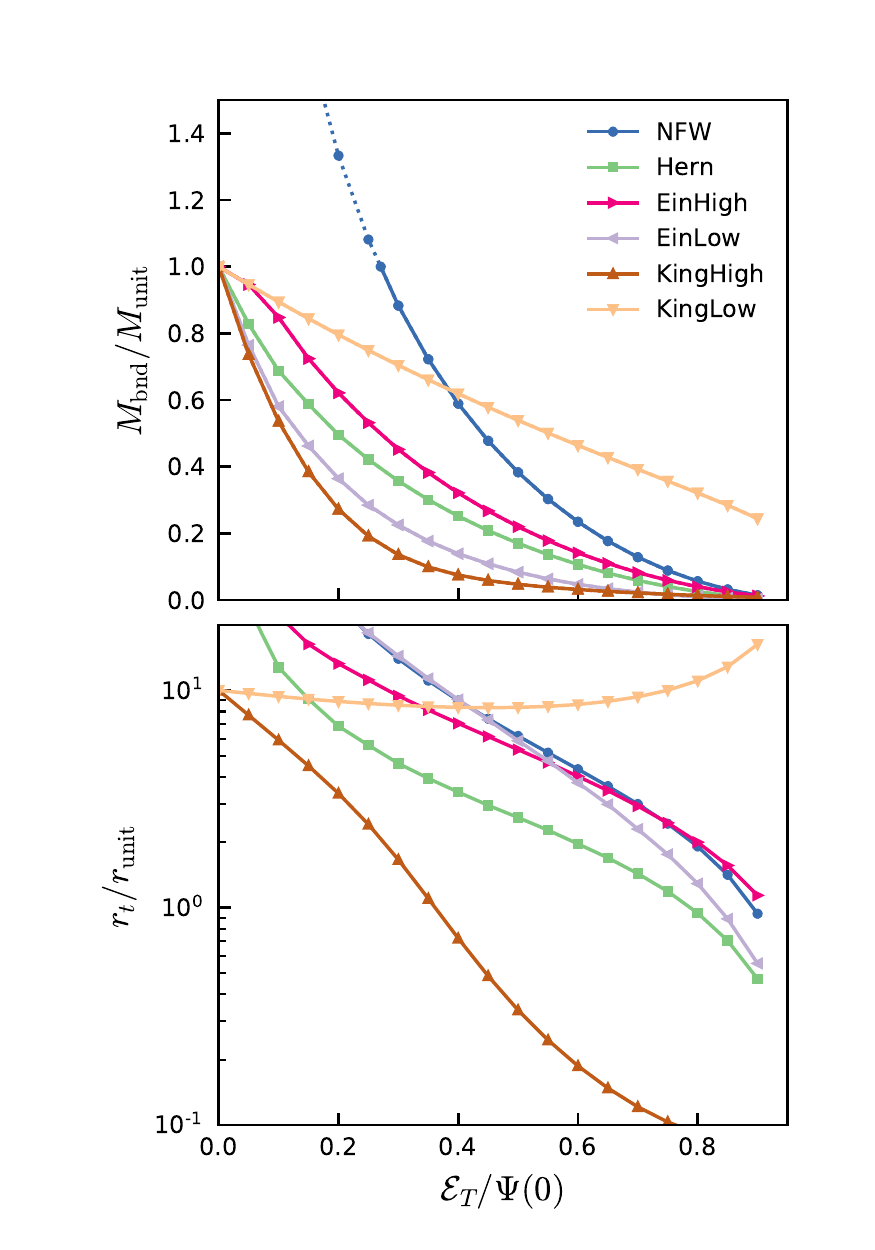}
	\caption{Relationship between the tidal energy $\mathcal{E}_T$, the bound mass of the satellite (top), and the tidal radius (bottom) for different initial satellite profiles. The relative energy, $\mathcal{E}$, has been normalized by the relative potential of the untruncated profile at $r=0$, $\Psi(0)$. Note that since the NFW profile has a divergent mass profile, we have chosen $M_{\rm unit}$ to correspond to an NFW profile truncated at $\mathcal{E}_{T}=0.27$; therefore, $M_{\rm bnd}/M_{\rm unit}$ can be greater than 1 (dotted line). In general, the bound mass and truncation radius decrease with tidal energy, with the exception of the KingLow profile. The KingLow profile is the only cored profile, and its truncation radius increases as it is tidally stripped.}
	\label{fig:Mbnd_vs_Et}
\end{figure}

Though the model presented in Section~\ref{sec:ourmodel_I} most naturally expresses the lowered density profiles in terms of a truncation energy, $\mathcal{E}_{T}$, the model can alternatively be parameterized by either the truncation radius, $r_{t}$, or the total bound mass of the satellite, $M_{\rm bnd}$. In Fig.~\ref{fig:Mbnd_vs_Et} we show the relationship between these three parameters. Generally, there is a monotonic relationship between the bound mass and the tidal energy or the tidal radius. An exception is the tidal radius for the KingLow models; interestingly in this case $r_{\rm t}$ \emph{increases} for large values of $\mathcal{E}_{\rm_T}/\Psi(0)$. We will show in Section~\ref{sec:truncprofs} that this prediction is also consistent with the simulation results.% (see Fig.~\ref{fig:ProfsFits}).

\begin{figure*}
	\includegraphics{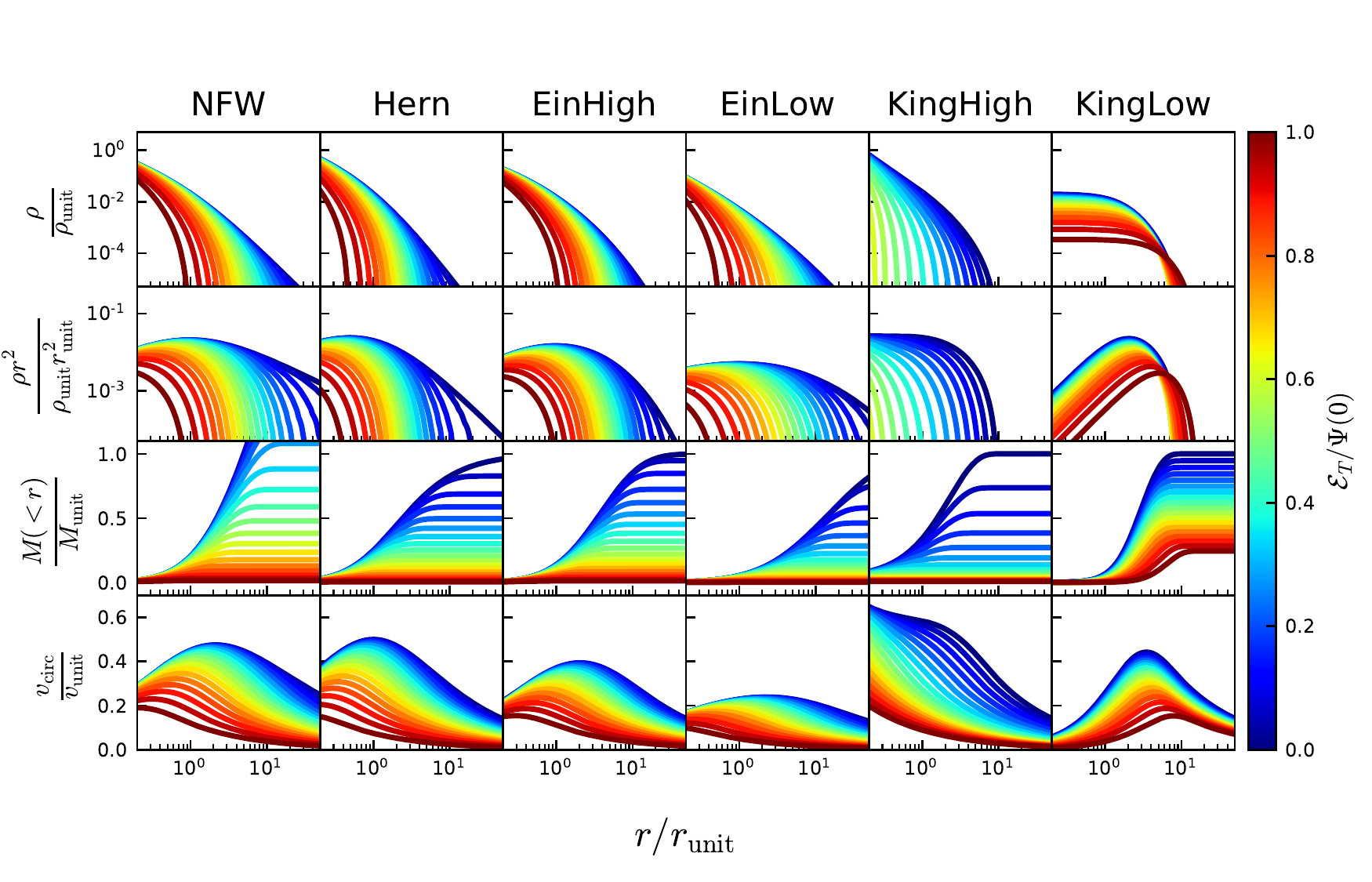}
	\caption{Predicted evolution for different initial density profiles (columns), assuming sharp truncation of the distribution function in energy. The EinLow and EinHigh profiles have $\alpha$ values of $\alpha=0.15$ and $0.3$, respectively, while KingHigh and KingLow profiles have central potentials of $P_0=12$ and $3$, respectively. The curves are coloured by the value of the tidal energy, $\mathcal{E}_{T}$, normalized by the relative potential of the untruncated profile at $r=0$, $\Psi(0)$.  The density profile goes to zero at the truncation radius, $r_t$. The relation between the tidal energy, truncation radius and bound mass is given in Fig.~\ref{fig:Mbnd_vs_Et}.}
	\label{fig:TruncProfs}
\end{figure*}

In Fig.~\ref{fig:TruncProfs} we show the truncated density profiles and various related quantities, coloured by the tidal energy, $\mathcal{E}_{\rm_T}$ normalized by the relative potential of the untruncated profile at $r=0$, $\Psi(0)$. For most of the halo models, as the satellites are stripped they decrease in density at all radii, though most of this decrease is at large radii. An exception is the KingLow models, where  $r_t$ increases, with a corresponding increase in density at larger radii. We note that when the density profile is cored (as in the centre of the King models), the central density decreases significantly as mass is lost, while for the cuspier profiles the central density is conserved. This is consistent with results from \cite{penarrubia2010}. 

We also find that the scale radius of the NFW, Hernquist and Einasto profiles (the point where the logarithmic slope of the density profile is $-2$, estimated from the location of the maximum of $\rho r^2$) decreases as the halo becomes tidally stripped. In principle, this could lead to an increase in the concentration parameter, though the virial radius is also decreasing; we examine halo concentrations further in a companion paper \citep[][in prep]{drakos2022}. Finally, the location and amplitude of the peak of the circular velocity curve both decrease with increasing $\mathcal{E}_T/\Psi(0)$ for all profiles except KingLow.

%%%%%%%%%%%%%%%%%%%%%%%%%%%%%%%%%%%%%%%%%%%%%%%%%%
%%%%%%%%%%%%%%%%%%%%%%%%%%%%%%%%%%%%%%%%%%%%%%%%%%
%%%%%%%%%%%%%%%%%%%%%%%%%%%%%%%%%%%%%%%%%%%%%%%%%%	
\section{Model comparison to simulations} \label{sec:univers} 
%%%%%%%%%%%%%%%%%%%%%%%%%%%%%%%%%%%%%%%%%%%%%%%%%%
%%%%%%%%%%%%%%%%%%%%%%%%%%%%%%%%%%%%%%%%%%%%%%%%%%
%%%%%%%%%%%%%%%%%%%%%%%%%%%%%%%%%%%%%%%%%%%%%%%%%%	

We have already demonstrated in Papers~I and II that the energy truncation approach provides an excellent description of the evolution of  systems with NFW profiles. The purpose of this section is to test whether this approach, and the mass loss model developed in Paper~II,  works as well for the other satellite models described in Section~\ref{sec:collis}. We start by comparing the predicted and observed mass-loss rates, before returning to the profile evolution.

%%%%%%%%%%%%%%%%%%%%%%%%%%%%%%%%%%%%%%%%%%%%%%%%%%
\subsection{Mass-loss rates}
\begin{figure*}
\includegraphics[]{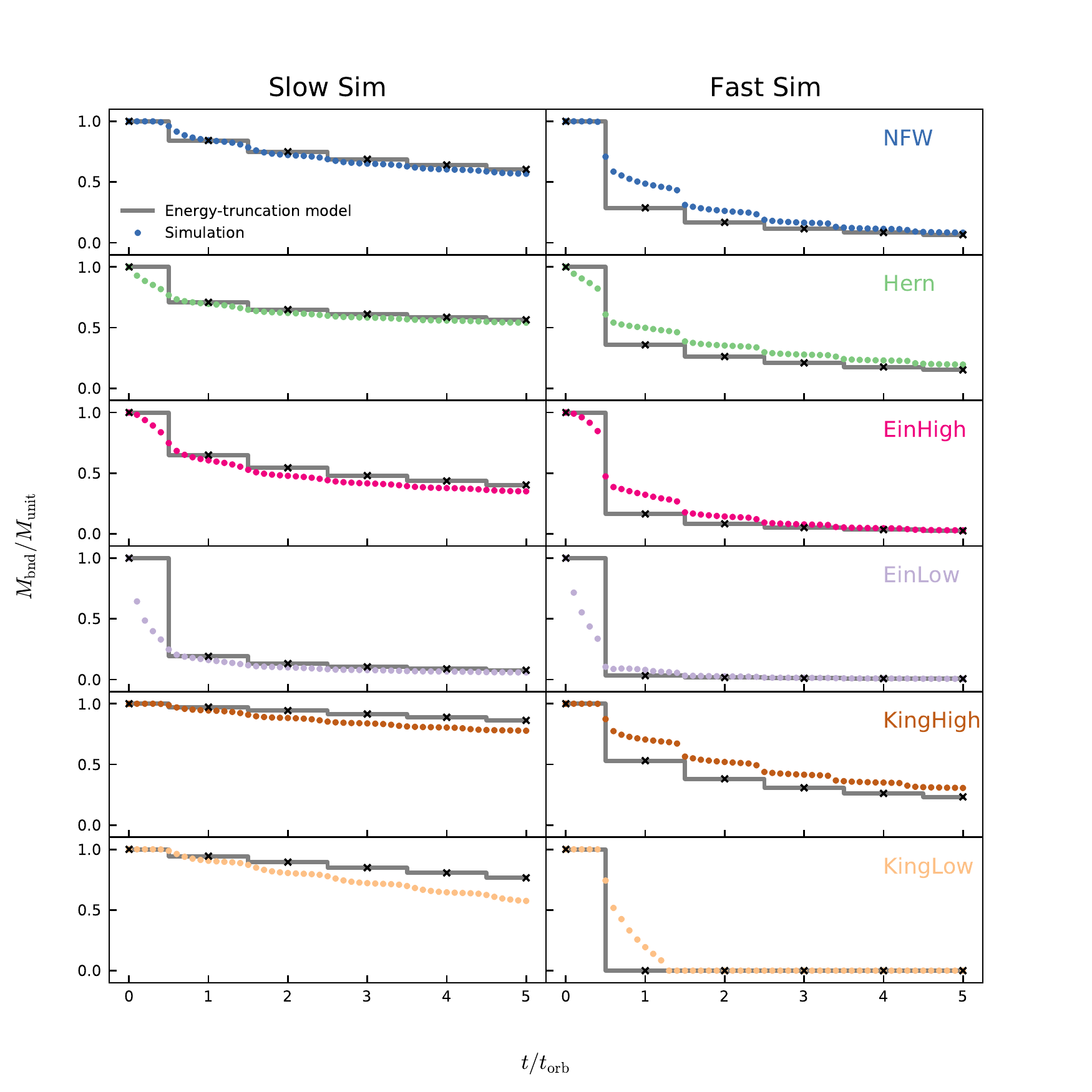}
\caption{Bound satellite mass as a function of time in the Slow and Fast Simulations, calculated as described in Section~\ref{sec:model} (points), compared to predictions of the mass loss from our model (grey lines). The black crosses correspond to the model prediction at apocenter. Each row corresponds to a different density profile, as indicated. In general, the energy-truncation model agrees with the simulation results to within 20 per cent or better.}
	\label{fig:MassLossCurves}
\end{figure*}

%%%%%%%%%%%%%%%%%%%%%%%%%%%%%%%%%%%%%%%%%%%%%%%%%%
In Fig.~\ref{fig:MassLossCurves} we compare the bound mass of the simulated satellites versus time, calculated as described in Section~\ref{sec:orbits}, to the predictions of the mass loss model summarized in Section~\ref{sec:ourmodel_III}. We assume the mass loss predicted by the model on each orbit is removed by the time the system reaches apocentre. With this convention for the Slow Simulation (left column), we get excellent agreement  (the bound mass is predicted to 5 per cent  or better accuracy over the first 5 orbits) for the NFW, Hern and EinLow profiles and good agreement (10 per cent or better accuracy over the first 5 orbits) for the EinHigh and KingHigh profiles. We get less accurate predictions ($\leq$20 per cent accuracy over the first 5 orbits) for the KingLow models. In all cases, mass loss is slightly faster in the simulations than predicted by the model.

For the Fast Simulations (right column), the observed mass-loss rates are slightly slower than predicted, particularly on the first few orbits, where the predictions for most models overestimate mass loss by $15$--$20$ per cent. After a few orbits, however, the simulations have caught up with the model predictions, and  we once again get good agreement ($\le$10 per cent accuracy) for all models. Additionally, the model correctly predicts the complete disruption of the KingLow profile by the second pericentric passage. 

While our model does a reasonable job of predicting the evolution of the mass distribution of the remnants, there are some discrepancies between the total mass-loss rates predicted by the model and the rates measured in the simulations. Defining which particles are bound to the simulated subhalo is not a straightforward procedure, however, and often includes particles that are temporarily bound but in the process of leaving the system \citep{penarrubia2009}. We show evidence below that marginally bound mass in the outer parts of the satellites may indeed affect the predicted mass-loss rates (see e.g. Fig.~\ref{fig:ProfsFits}). 

The mass-loss picture is further complicated by the fact that it is often difficult to determine whether mass loss in simulations is due to physical processes or numerical artefacts \citep[e.g.][]{vandenbosch2018, vandenbosch2018b}. Additionally, our mass loss model does not include dynamical friction, which will cause the pericentric radius to decrease in time. While dynamical friction can largely be neglected, as the host halo is modelled as a smooth potential, there is some self dynamical friction, from the remnant orbiting through tidal debris \citep[e.g.][]{miller2020}.

%%%%%%%%%%%%%%%%%%%%%%%%%%%%%%%%%%%%%%%%%%%%%%%%%%
\subsection{Profile evolution}
%%%%%%%%%%%%%%%%%%%%%%%%%%%%%%%%%%%%%%%%%%%%%%%%%%

\begin{figure*}
	\centering
	\subfloat{{\includegraphics[trim={0 0cm 0 1cm},clip=true]{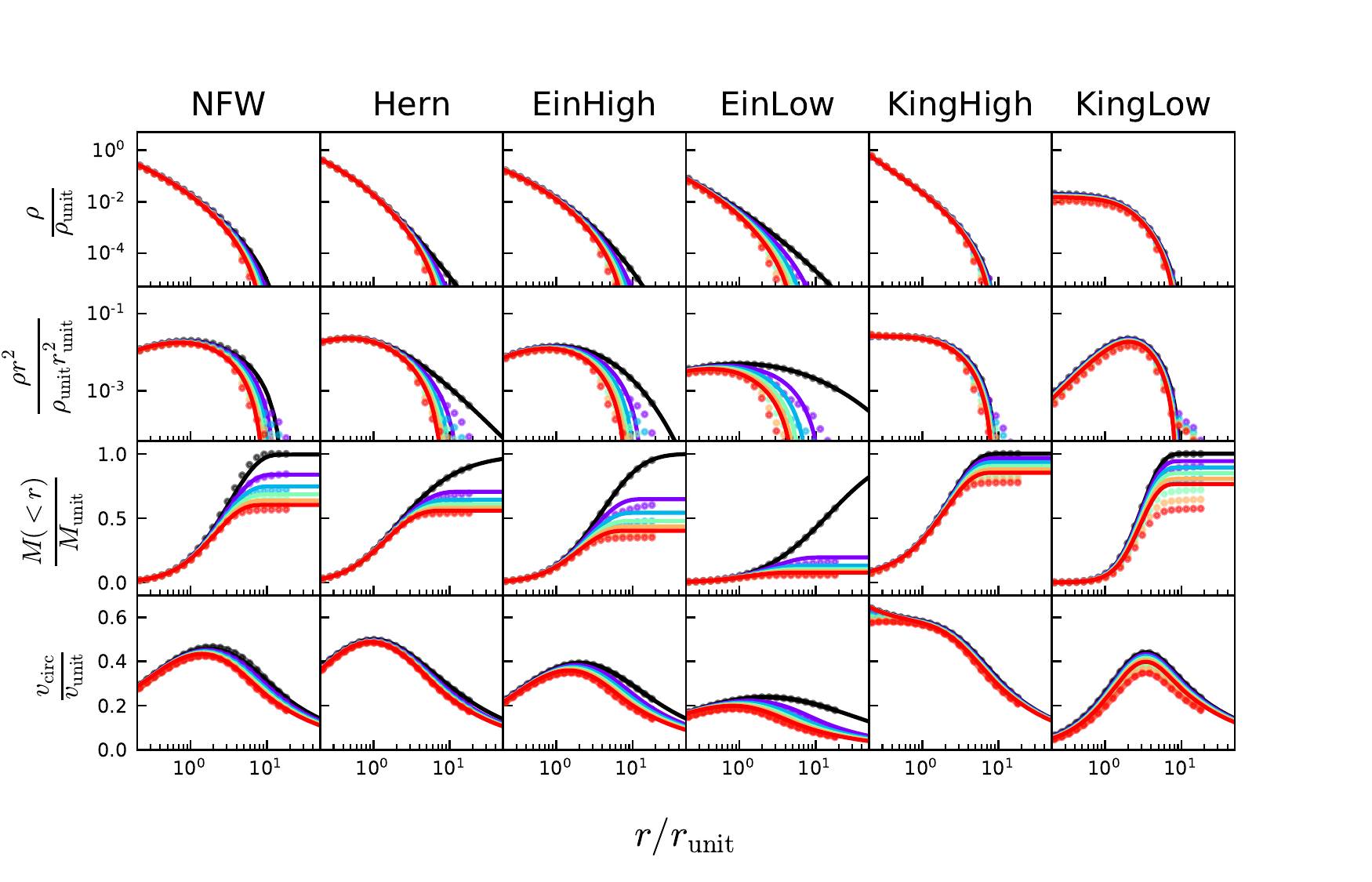}}}%
	
	\subfloat{{\includegraphics[trim={0 0 0 1cm},clip=true]{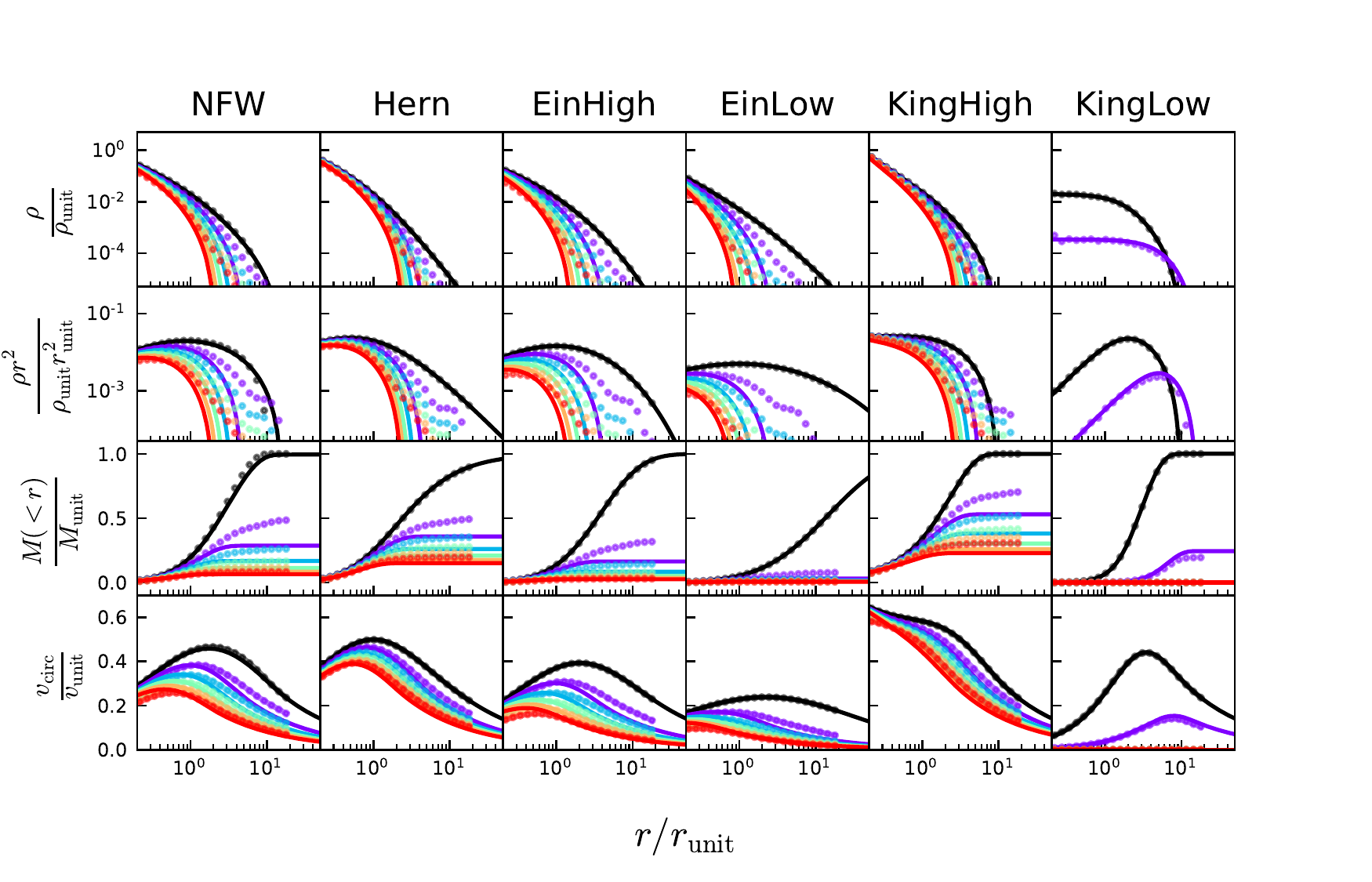}}}%
	\caption{
		Comparison of simulation results (points) and the energy-truncation model (lines) for the Slow (top) and Fast (bottom) Simulations. Different curves correspond to successive apocentric passages. The energy-truncation model agrees well with simulations, except close to the truncation radius where some additional mass appears to be temporarily bound in the simulations. Note that for the cored KingLow profile (last column, bottom panel), our model correctly predicts the observed drop in the central density. Residuals for the density and mass profiles are shown in Appendix~\ref{sec:resid}.
	}
	\label{fig:ProfsFits}
\end{figure*}

In addition to predicting mass loss, our model also predicts the tidal energy on each orbit, and thus the full structural evolution of the satellite. In Fig.~\ref{fig:ProfsFits}, we show how the density, mass and circular velocity profiles of the simulations compare to the model predictions, for the Slow (top panels) and the Fast (bottom panels) simulations. For the Slow Simulations, the agreement is excellent, with the only significant discrepancy being at large radii for the EinLow profile. 

In the Fast Simulations, the agreement is also excellent at small radii, with the exception of a slight disagreement in the circular velocity curves of the KingHigh profiles at small radii; we suspect this is due to numerical relaxation in the simulation. There are more significant discrepancies at large radii for most of the profiles. The density profiles measured in the simulations drop abruptly close to the predicted tidal radius, but then they extend well beyond this radius with a shallower slope. As discussed in Paper~II and in \cite{penarrubia2009}, this part of the profile  includes transitional material that is still bound but moving outwards, and will be lost mostly on the next orbit. Refining the procedure we use to identify bound particles in the simulation might resolve these discrepancies, and improve the agreement between predicted and observed mass-loss rates in the case of orbits with rapid mass loss. It may also be the case the energy-truncation fails to accurately predict the evolution of the outskirts of the profile, due to the equilibrium assumptions inherent in the model. Understanding the detailed mass-loss and time scales is an interesting question, and one we leave to future work.

Overall, our model appears to be universally valid for a wide range of density profiles. The predictions for the mass loss rate are generally very accurate, and especially so for the models that might describe dark matter structure (NFW, Hernquist and Einasto), predicting the reminding bound mass to within 5-10 per cent or better in most cases over the first 5 orbits. The predictions for the structure of the bound remnants are accurate except in the outer regions close to the tidal radius; the latter discrepancy might be resolved by adding a timescale for mass loss to our model, or refining our definition of bound mass in the simulations. The success of the  energy-truncation model is particularly impressive given that it has no free parameters to adjust.

%%%%%%%%%%%%%%%%%%%%%%%%%%%%%%%%%%%%%%%%%%%%%%%%%%	
\section{Central density predictions} \label{sec:compare}
%%%%%%%%%%%%%%%%%%%%%%%%%%%%%%%%%%%%%%%%%%%%%%%%%%	

\begin{figure*}
	\includegraphics{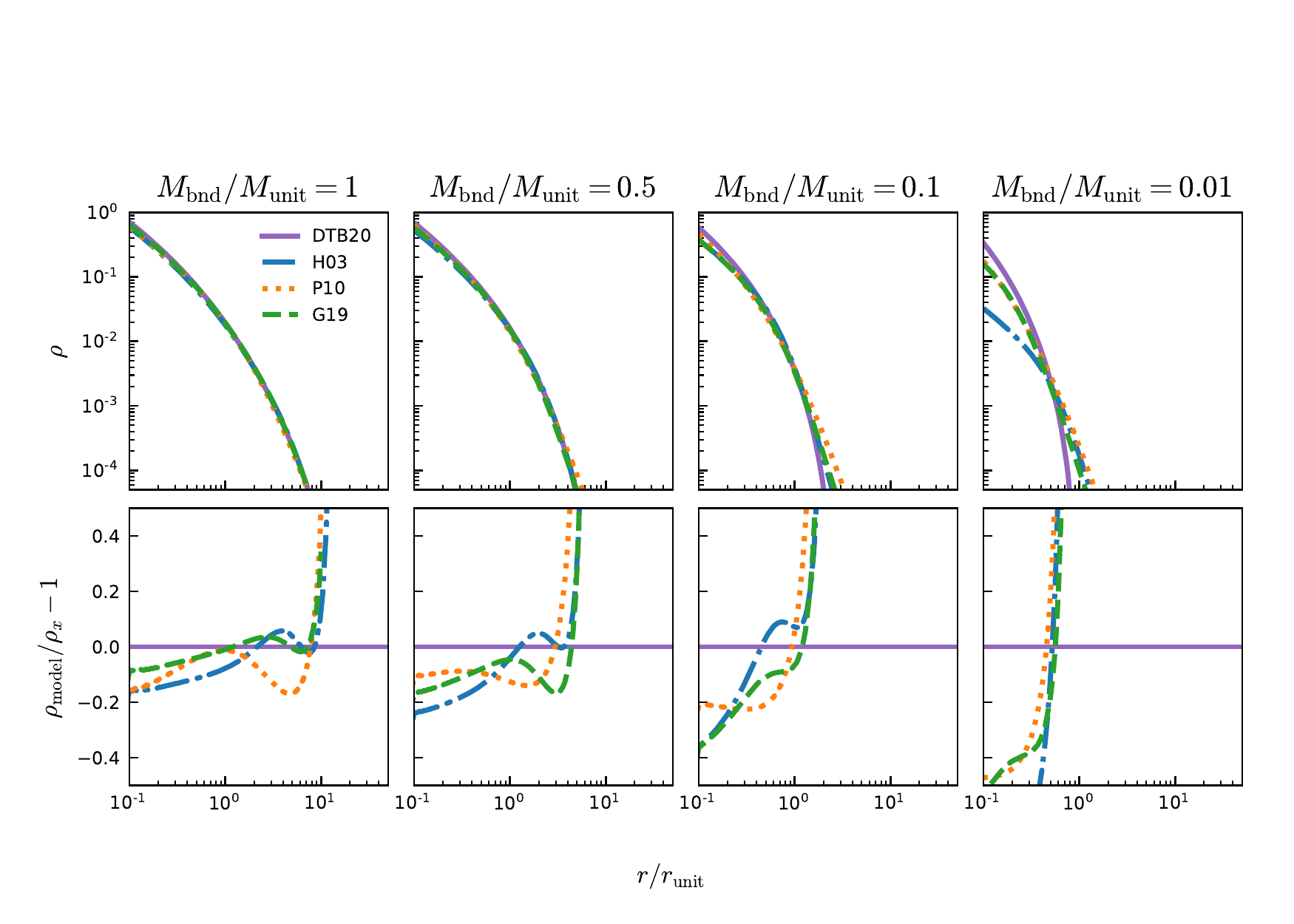}
	\caption{Our model prediction (DTB20) 
	for the density profile of a tidally stripped NFW system, compared to previous empirical models of the stripped profile, including \citet{hayashi2003} (H03), \citet{penarrubia2010} (P10) and \citet{green2019} (G19). Each column is a different bound mass fraction. The bottom panels show the residuals, calculated as $\rho_{\rm model}/\rho_x -1$, where $\rho_{\rm model}$ is our model from Paper I, and $\rho_x$ are the empirical models. Our model predicts larger central densities and slightly sharper truncation compared to the others, particularly at low bound mass fractions.}
	\label{fig:NFW_compare}
\end{figure*}

The main difference between the energy-truncation model and previous empirical models of mass loss in the literature is that it predicts a conserved central density in the tidally stripped systems, even for small bound mass fractions. This is illustrated in Fig.~\ref{fig:NFW_compare}, where we compare our NFW model predictions to empirical models of stripped NFW systems from \cite{hayashi2003}, \cite{penarrubia2010} and \cite{green2019}.\footnote{Note that the bound mass fraction is defined in these papers as $M/M_{\rm NFW}(<r_{\rm cut})$. In the following discussion, we convert all bound mass fractions to  $M/M_{\rm unit}$.} As shown in Paper II, the recent model from \cite{green2019} is likely the most accurate description of tidally stripped NFW haloes from single-subhalo simulations, while the model from \cite{hayashi2003} underestimates the central density.

While the empirical models capture by design the results of the underlying idealized simulations, it is unclear whether these simulations accurately describe realistic cosmological situations. Empirical fits to simulations will also fit numerical artefacts present in the results. For instance, \cite{hayashi2003} used an approximation in their initial conditions which caused a drop in the central density in their simulated haloes that was numerical in nature; this in turn is reflected in their empirical model for the stripped profile \citep{kazantzidis2004}.   While the more recent paper by \cite{green2019}  uses simulations carefully calibrated to understand the effects of numerical relaxation on mass loss \citep{ogiya2019}, it is still unclear how the very central regions of the haloes are affected. Indeed, separate work from \cite{errani2020} supports the idea that centrally-divergent cusps are never disrupted. 

Following \cite{binney}, given the number of particles within radius $r$, $N(<r)$, each of mass $m$, we can calculate a relaxation time scale, 
\begin{equation}
t_{\rm rel}(r) \approx 0.1 \frac{\sqrt{N(<r)}}{\ln N(<r)} \sqrt{\frac{r^{3}}{G m}} \,\,\, ,
\end{equation}
which is the time it takes for a typical particle's velocity to change by an order of itself, and an evaporation time scale, $t_{\rm evap}(r) \approx 136 \,t_{\rm rel}(r)$, which is the time for a typical particle to reach escape speed. Since both times increase with radius, given a time, $t$, we can then calculate relaxation and evaporation radii such that $t = t_{\rm rel}(r_{\rm rel}) $ and $t = t_{\rm evap}(r_{\rm evap}) $, respectively. This will give us an indication of the radii inside which resolution effects might dominate the evolution. Note that in Fig.~\ref{fig:NFW_compare}, we cannot determine at which radius numerical relaxation and evaporation will dominate, since there is no temporal information; the time for a subhalo to reach the indicated mass fraction would depend on the specific orbit and background potential considered.

The slow mass loss case is the one in which the assumptions of the energy-truncation model appear most valid (Paper II), such that we might expect the central density to be conserved. The relative relaxation rate increases as the number of particles in a system decreases, however, and we are unable to evolve subhaloes to very low masses for the Slow Simulation and still have enough particles to preserve the subhalo against numerical effects. Thus, we cannot reliably measure the central density of the halo at late times in the slow mass-loss simulations. 

\begin{figure*}
	\includegraphics{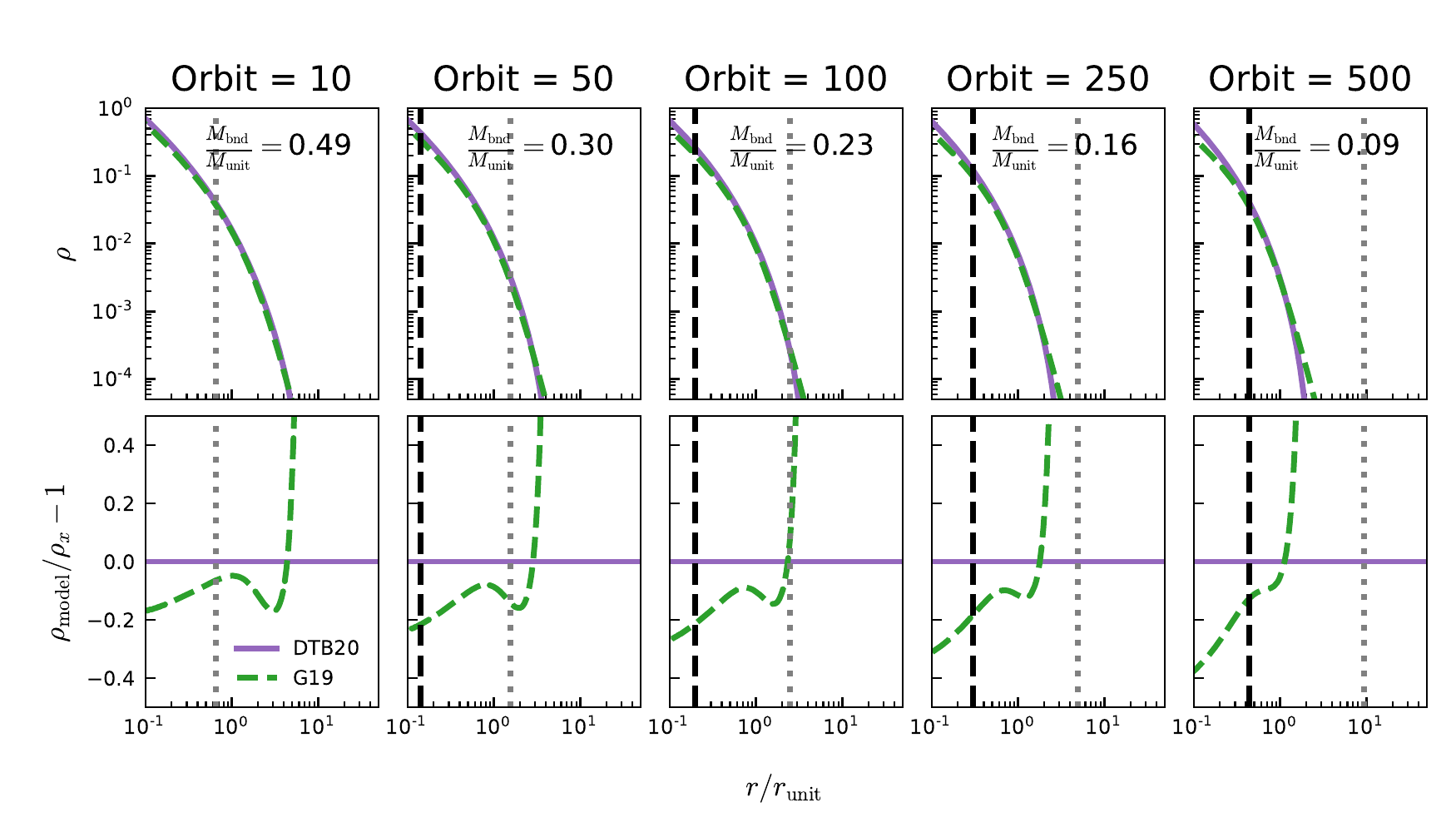}
	\caption{
	The evolution of the density profile, as in Fig.~\ref{fig:NFW_compare}, but comparing our model with the model of \citet{green2019}, after the number of orbits labelled. The dotted grey lines and dashed black lines show the relaxation radius and evaporation radius, respectively. The difference in central density between our model and the model of \citet{green2019} generally lies at radii unresolved in the simulations.
		}
	\label{fig:NFW_compare_relax}
\end{figure*}

We can, however, use our analytic mass-loss model, as outlined in Section~\ref{sec:model}, to predict the density profile after 10, 50, 100, 250 and 500 orbits. Fig.~\ref{fig:NFW_compare_relax} shows this prediction for the orbit of the Slow Simulation. Additionally, we show the profile predicted from \cite{green2019} (using the same enclosed mass), and the relaxation and evaporation radii (vertical lines). The simulations used to calibrate the model in \cite{green2019} were run at a similar resolution to ours ($\sim 10^6$ particles), and therefore should have similar resolution effects.  With the exception of the density at large radii (which depends on the details of how bound mass is calculated), the results of \cite{green2019} are nearly indistinguishable from our model for $r>r_{\rm evap}$. They differ significantly from the model predictions at smaller radii, but this is precisely where we expect relaxation to affect the results. 

In summary, while we cannot resolve the question of central density conservation definitively with our own simulations, we have shown the previous predictions in the innermost parts of the profile need to be treated with caution, and that the question is not yet resolved by previous work in the literature. In a companion paper \citep[][in prep]{drakos2022}, we explore the sensitivity of dark matter annihilation rates and galaxy lensing signals to the assumed behaviour of the innermost part of the density profile.

%%%%%%%%%%%%%%%%%%%%%%%%%%%%%%%%%%%%%%%%%%%%%%%%%%	
%%%%%%%%%%%%%%%%%%%%%%%%%%%%%%%%%%%%%%%%%%%%%%%%%%	
\section{Physical interpretation} \label{sec:physical}
%%%%%%%%%%%%%%%%%%%%%%%%%%%%%%%%%%%%%%%%%%%%%%%%%%	
%%%%%%%%%%%%%%%%%%%%%%%%%%%%%%%%%%%%%%%%%%%%%%%%%%	

In the previous sections, we demonstrated the overall accuracy of the energy-truncation model for a broad range of profiles. In this section, will consider the physical justification for this model in more detail. As demonstrated in \cite{choi2009} and Paper II, during tidal stripping, mass loss appears to be ordered primarily by the original relative energy of the particles. The energy-truncation model captures this trend by shifting the relative energy of all particles by a constant amount over each orbit, and unbinding those under some threshold binding energy. However it does not explicitly state why the ordering in relative energy should be conserved, or equivalently, why the change in relative energy is approximately constant. 

In general, over the course of an orbit, a particle can become unbound from a subhalo because (1) the velocity of the particle changes, e.g.~due to a kick from a rapid perturbation or, (2) the potential of the system changes, due to a change in the external field, or due to mass loss from the subhalo itself. When estimating the relative contributions of these effects, we have to be clear which frame is used to calculate kinetic energies, and which mass is included in the calculation the potential. In this section, we will always define particle velocities $v$ and radii $r$ in the frame of the satellite, and will we calculate the satellite potential from the self-bound mass defined previously. 

Given these conventions, we can write the two contributions to the change in energy as
\begin{equation}
\Delta \mathcal{E} = \Delta \mathcal{E}_K  + \Delta \mathcal{E}_P \,\,\, ,
\end{equation}
where
\begin{equation} \label{eq:deltaE_K}
\Delta \mathcal{E}_K =-\left[ \dfrac{1}{2} (\mathbf{v_0} + \Delta \mathbf{v})^2 -\dfrac{1}{2} \mathbf{v}^2\right]=- \mathbf{v} \cdot \Delta \mathbf{v}- \dfrac{1}{2} (\Delta \mathbf{v})^2\, ,
\end{equation}
assuming the velocity of the particle has changed from $\mathbf{v}$ to $\mathbf{v} + \Delta \mathbf{v}$.

If we make the impulse approximation --- assuming the satellite undergoes a perturbation or `shock' much shorter than most other timescales in the problem, so particle positions are approximately constant throughout the event \citep[e.g.][]{taylor2001a} --- we can get an estimate of the energy change from the shock by treating it as a point-mass perturber $M_p$, passing by the satellite on a linear trajectory with a velocity $v_p$ and an impact parameter $b$. Tidal heating from the shock should accelerate particles, imparting a change in velocity
\begin{equation}
\Delta \mathbf{v} = \dfrac{2 G M_p}{v_p b^2} [-x,y,0]\,\,\, ,
\end{equation}
where $(x,y)$ is the position of the particle in the plane of the encounter \citep[see, e.g.][]{MVW}. Since $\Delta \mathbf{v} \sim \mathbf{r}$, $\mathbf{v}^2 \sim r^2$, and the impulse approximation predicts that $|\Delta \mathcal{E}_K|\propto r^2$ at large radii, and that it may also have some additional energy or velocity dependence from the first term on the right-hand-side of Equation~\eqref{eq:deltaE_K}.
On the other hand, at small radii, adiabatic shielding prevents particles from experiencing much net acceleration, since their internal orbital timescales become shorter than the timescale of the changing tidal field close to the centre of the satellite \citep[e.g.][]{gnedin1999}. Thus the change in $\mathcal{E}_K$ should be approximately zero in this limit. 
After this instantaneous change in kinetic energy, the system re-virializes. The negative heat capacity of the system suggests that there will be a resulting decrease in kinetic energy (an increase in $\Delta\mathcal{E}_K$).

%If the satellite is in virial equilibrium before and after the encounter, this would imply  $\Delta_\mathcal{E} = \Delta\mathcal{E}_K = - \Delta\mathcal{E}_P /2 $. Since the system is not isolated, these relations do not old perfectly bound particles actually have.... 

\begin{figure*}
	\includegraphics[]{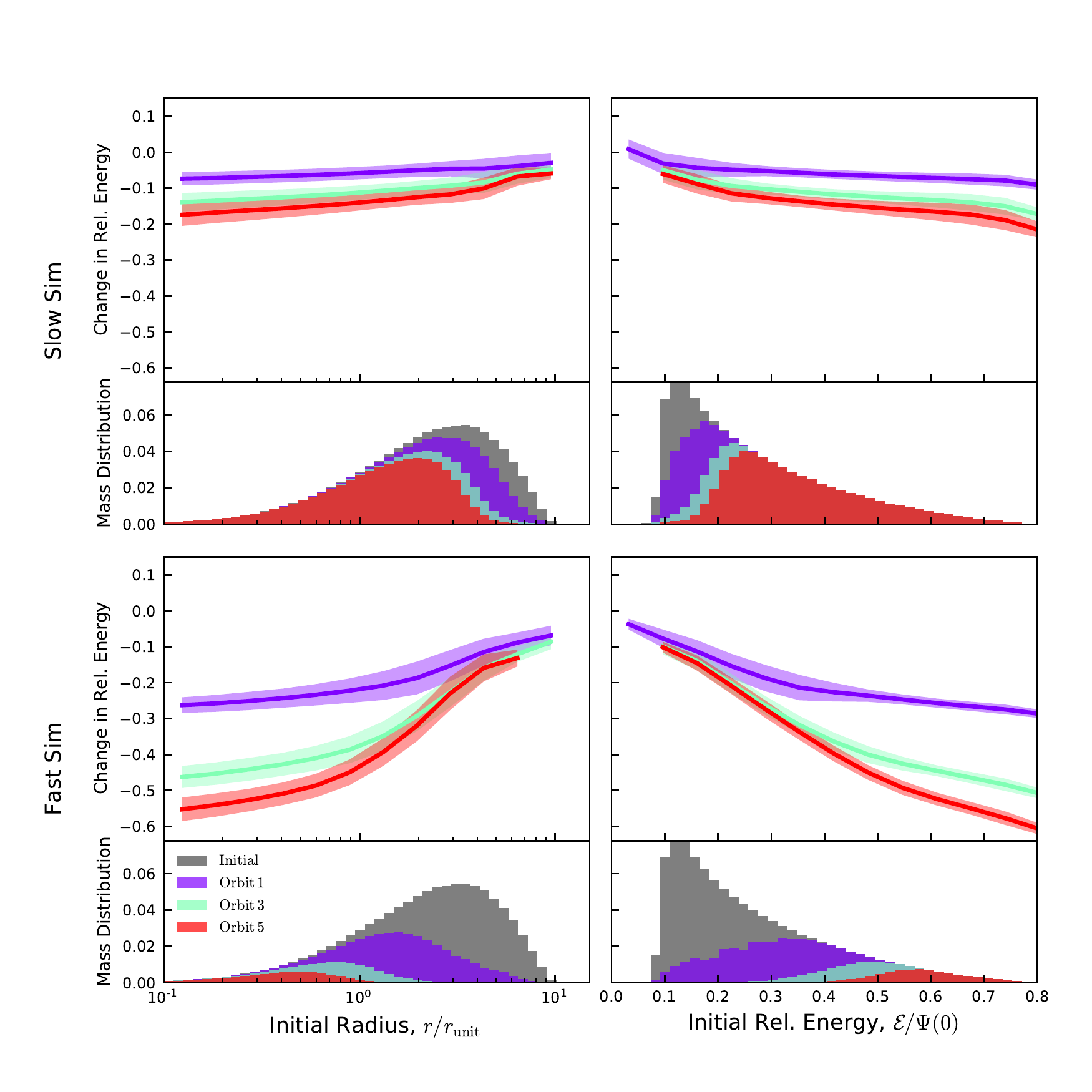}
	\caption{
	The average change in relative energy of the satellite self-bound particles ($\Delta \mathcal{E} $). Note that negative changes in energies correspond to particles becoming less bound to the satellite. Specific energies are calculated for both the Slow Simulation (top) and Fast Simulation (bottom) after one, three and five orbits. Shaded regions show the $1-\sigma$ standard deviation in each bin. We also show the mass distribution of particles as a function of initial radius (left) and initial energy (right).
	}
	\label{fig:DeltaE}
\end{figure*}

Fig.~\ref{fig:DeltaE} shows the change in the specific energy of the bound particles over 5 orbits. The top panels show the change in relative energy as a function of initial radius (left-hand) and initial energy (right-hand panels), while the bottom panels show how the mass is distributed in the system, as a function of the same variables. While there some variation in $|\Delta \mathcal{E}|$ across the satellite (for particles that remain bound to the system, $|\Delta \mathcal{E}|$  decreases with radius and also with binding energy), overall it is fairly constant, varying by $\sim 10$ per cent over the range of radius or energy that contains most of the mass\footnote{Note that in Paper II, we showed that the scatter in the initial relative energy of particles stripped on a given orbit was also $\sim 10$ per cent -- see Fig.~5 of Paper II.}. We see larger variations only at large radii or low binding energies, where particles are close to being stripped. Overall, the mean shift in relative energy from orbit to orbit is fairly constant as a function of particle radius or initial energy, and also from particle to particle, even after 5 orbits. This shows why the energy truncation model produces a good approximation to the rate and radial dependence of the actual mass loss, particularly in the slow mass-loss case. The largest variations in $|\Delta \mathcal{E}|$ are close to the edge of the system, where particles are close to being stripped.

The weak and declining dependence of $|\Delta \mathcal{E}|$ on particle radius may seem counter-intuitive, since the particles at large radii are expected to get larger velocity kicks in the satellite frame. Overall, the change in relative energy does not scale with radius as expected for at least two reasons. First, particles at small radii tend to be more bound, and therefore can experience more acceleration and still remain bound, while particles at large radii that receive large kicks will be lost from the system. We do not track this unbound material in our model. This mass loss reduces the average energy change in the material that remains bound, and may also explain why the energy change decreases in magnitude quite rapidly close to the boundary of the system.
The second reason is that the main contribution to the relative energy change is actually from the potential term, which decreases more or less uniformly over the whole inner part of the satellite, as mass is stripped primarily from the outside in. This term will deviate from a constant only in the outer part of the satellite where some but not all of the mass is lost, so here too this contributes to the rapid change in $|\Delta \mathcal{E}|$ at large radii.

\begin{figure*}
	\includegraphics[]{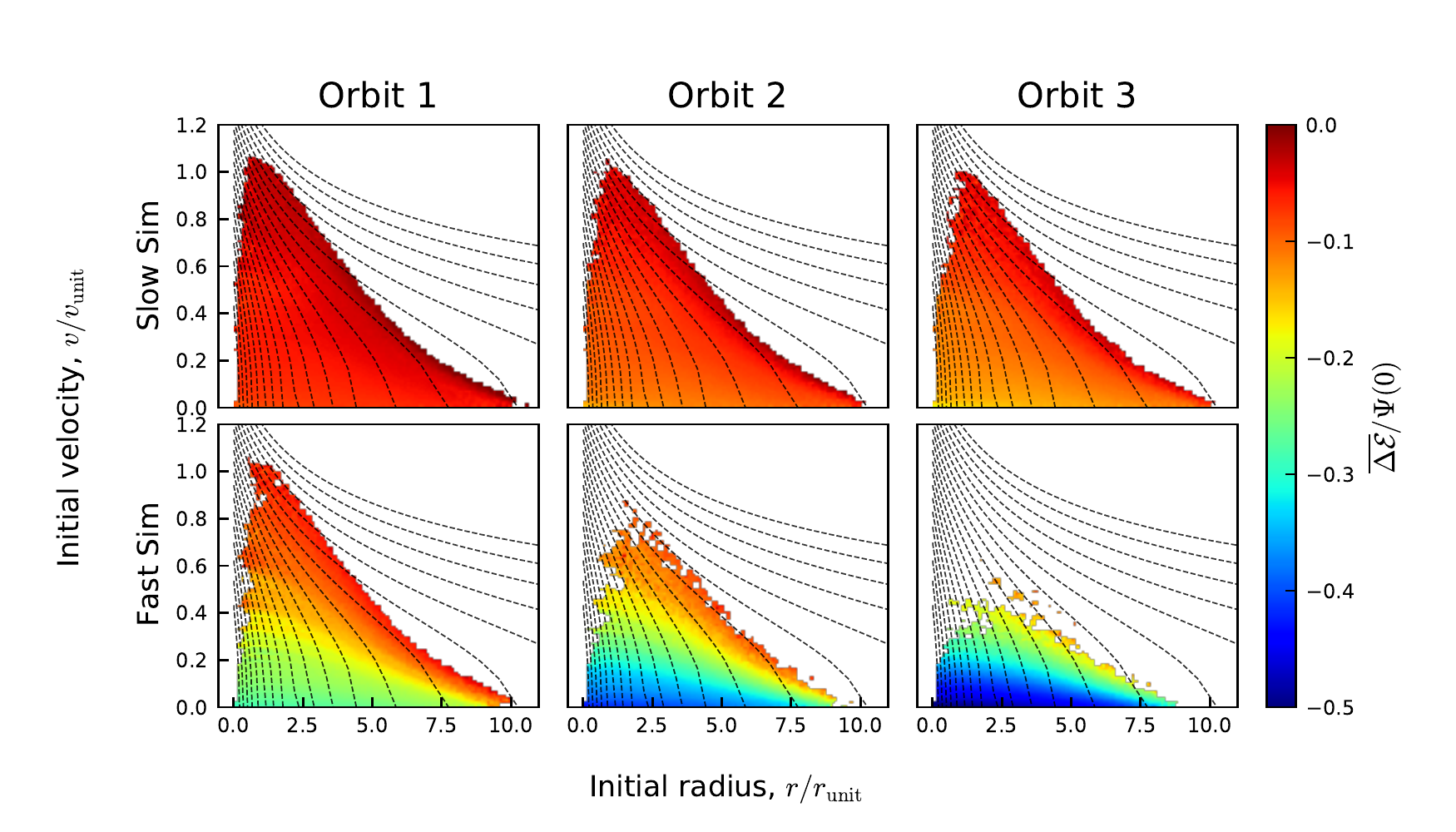}
	\caption{
The average particle change in relative energy, $\overline{ \Delta \mathcal{E} }$, as a function of the initial location in phase space for the self-bound remnant. We plot contours of constant energy in an NFW profile with dashed lines. The change of energy is constant along the edge of the system, which roughly corresponds to a constant energy,  indicating that change in relative energy is primarily a function of particles initial energy. The decrease in $\mathcal{E}$ in the center of the system (initial $(r,v)\approx (0,0)$) is due to the decreased self-bound mass of the satellite.}
	\label{fig:PhasePlots}
\end{figure*}

We can also consider which particles are removed in phase space. In Fig.~\ref{fig:PhasePlots} we show the average change in relative energy as a function of initial particle position. Close to the phase-space boundary of the system (at a constant relative energy), the change in relative energy is constant. The decrease in relative energy in the centre of the system is due to the overall mass of the system decreasing, causing particles to become less bound. This figure demonstrates that $\Delta \mathcal{E}_P$ for the self-bound particles mainly depends on their initial energy, especially for the Slow Simulation.   

In summary, there are several reasons 
why the energy-truncation model provides a good estimate of the rate and radial dependence of mass loss, or equivalently, why mass loss remains ordered by the original relative energy to a significant degree. While simple arguments predict that tidal heating should depend on particle radius and possibly also on energy, this variation is reduced by selective mass loss at large radii and/or low energies, and the net variation in the remaining bound mass is much smaller. Furthermore, the total change in relative energy is actually dominated by the potential energy change, which is relatively constant throughout the inner part of the satellite.

Overall, the physical picture for mass-loss in our framework is that as the satellite moves through the background potential, the tidal fields heat its particle orbits and deform its self-potential (see Appendix~\ref{sec:boost}). The lowest $\mathcal{E}$ particles rapidly escape, since they are no longer bound to the satellite. Once this mass loss has occurred, however, the potential well of the system decreases by a roughly constant amount, and the energies of all the particles are consequently shifted while remaining ordered. Our model approximates this process by assuming the initial distribution function is lowered by a constant amount on each orbit. As discussed in Section~\ref{sec:conc}, future work will focus on improvements to this energy-truncation model, for instance by adding scatter to the change in relative energy.

%%%%%%%%%%%%%%%%%%%%%%%%%%%%%%%%%%%%%%%%%%%%%%%%%%
%%%%%%%%%%%%%%%%%%%%%%%%%%%%%%%%%%%%%%%%%%%%%%%%%%
%%%%%%%%%%%%%%%%%%%%%%%%%%%%%%%%%%%%%%%%%%%%%%%%%%		
\section{Discussion} \label{sec:conc}
%%%%%%%%%%%%%%%%%%%%%%%%%%%%%%%%%%%%%%%%%%%%%%%%%%
%%%%%%%%%%%%%%%%%%%%%%%%%%%%%%%%%%%%%%%%%%%%%%%%%%
%%%%%%%%%%%%%%%%%%%%%%%%%%%%%%%%%%%%%%%%%%%%%%%%%%		

%Summary
In Paper~I, we described an approach to modelling tidal mass loss based on truncating and lowering the subhalo DF by a specified tidal energy. In Paper~II we used this approach to develop a model for mass loss with zero free parameters, and showed that it provided a good description of the evolution of subhaloes with NFW density profiles. In this work, we have demonstrated that energy truncation and our mass loss model can be applied to a wide range of other density profiles, with similar accuracy. We found the model does an excellent job of describing the density, mass and velocity profile evolution in all the tested cases. Additionally, our model naturally captures the effect that cuspy profiles conserve central density, while cored profiles have a large decrease in central density, as described in \cite{penarrubia2010}.

%Initial profile

Idealized simulations of individual subhaloes (such as the ones used in this paper) require an assumption about the initial profile; this is typically taken to be NFW, despite evidence that dark matter haloes are probably better described by Einasto profiles \citep[e.g.][]{navarro2010,klypin2016}.
Additionally, a recent study by \cite{brown2020} indicates that the universal profile might be due to the narrow range of initial conditions used in simulations, and the diversity of dark matter profiles should be much larger than previously thought.
Regardless, the density profile of dark matter haloes in cosmological $N$-body simulations is never resolved below the scale of the softening length. To establish the true behaviour of haloes in their innermost regions, we will likely need a theoretical model for the origin of the universal density profile.

%Artificial disruption.. 
As discussed in Paper~II,  mass loss predictions are sensitive to the central density of the system. Artificial disruption is likely a huge problem in cosmological simulations on subhalo scales \citep{vandenbosch2018,errani2020,errani2021,green2021} and is thought to be caused by artificial constant-density cores on the scale of the resolution limit of the simulation that drive enhanced tidal disruption. Controlled simulations suggest that centerally-divergent profiles can never be fully disrupted by a smooth potential \citep[e.g.][]{kazantzidis2004,errani2020}. Similarly, \cite{amorisco2021} recently argued that the conditions of (1) a centrally divergent density profile $\rho \sim r^{-1}$ and (2) an isotropic phase space distribution are sufficient  to ensure that subhaloes can never be disrupted. These finding are in agreement with the energy-truncation model, which predicts that systems will naturally always retain a bound remnant as long as $\bar{\rho}(r) \to \infty$ as $r \to 0$.

%Early haloes
These issues are particularly interesting in the context of microhaloes. These structures form early in the universe, and their size is determined by the free streaming scale of the dark matter particles. 
Assuming dark matter particles have a mass of 100 GeV, the mass of the smallest microhaloes is approximately an Earth mass \citep[e.g.][]{diamanti2015}.
%It is expected that the central density profile of these microhaloes is steeper than that found in larger haloes, and thus they may contribute greatly to the dark matter annihilation signal. 
%Concentration
%Since microhaloes should have lower Einasto $\alpha$ values, the results in this paper from the EinLow Simulations are particularly pertinent to this discussion. We found the EinLow simulations have the largest change in boost factor compared to truncating the profile abruptly, and also are the only profile that show an increase in concentration as they are stripped. Interestingly, in \cite{drakos2019b}, where we studied the effect of major mergers on halo properties, we found that Einasto profiles with low $\alpha$ values were the only ones that decreased in concentration after major mergers. These results all suggest that profiles with cuspier centres (as found in the earliest haloes) evolve differently.
%Central density of haloes
These early haloes probably have cuspy profiles, with an inner slope of $\sim -1.5$ \citep[e.g.][]{ogiya2016}. Most evidence suggests that the steep central cusps of microhaloes are not conserved, and mergers are thought to drive profiles towards the universal density profile, with an inner slope of  $-1$. For instance, \cite{ishiyama2014} show that the cusp slope gradually becomes shallower with increasing halo mass. More recently, by using  cosmological zoom-in simulations,  \cite{wang2019} were able to simulate Earth-mass haloes at redshift zero, and find that they look NFW (with an inner slope of $-1$). 

Generally, it is thought that major mergers are responsible for this flattening \citep[e.g.][]{ogiya2016,angulo2017,delos2019a}. Contrary to this picture, however, isolated simulations of major mergers show that the inner densities of haloes are typically conserved \citep[e.g.][]{ kazantzidis2006,drakos2019a,drakos2019b}. A possible solution to this discrepancy is that isolated binary merger simulations are overly simplistic, and do not take into account the complex gravitational interactions to which haloes are subject. However, it is also possible that high central densities are not conserved in cosmological simulations due to numerical artefacts. %Therefore, it is unclear whether halo central densities change over time, and what  mechanisms could potentially be responsible for a decrease in central density.
The central densities of these microhaloes are especially important to dark matter annihilation calculations. For instance, 
\cite{ishiyama2014} found that microhaloes had higher densities than expected from extrapolations of the low-redshift concentration--mass--redshift relation, while \cite{okoli2018} showed that if these densities are conserved, it could increase estimates of the dark matter annihilation boost factor by up to two orders of magnitude. 
%Overall, \cite{okoli2018} find that concentrations of the smallest haloes are still uncertain by a factors of $\sim 5$ when extrapolated to low redshift. 

Overall, there is a growing body of evidence that the evolution of the inner central density of haloes and their concentrations are not well understood. This is particularly concerning since models of substructure evolution are key ingredients in lensing predictions and dark matter annihilation constraints. The consequences of this are examined in a companion paper, where we show that while lensing constraints are not particularly sensitive to assumptions about subhalo central density, dark matter annihilation signals depend very sensitively on this quantity, and derived constraints on dark matter properties are therefore still  uncertain 
 \citep[][in prep]{drakos2022}.

%Model Simplifications
The energy-truncation model used in this work is a promising approach to modelling substructure evolution. As examined thoroughly in Paper II, the energy truncation model includes a number of simplifying assumptions which could be relaxed to make more accurate predictions. A straightforward extension of the model would be allow some additional particle-to-particle scatter in the energy change between orbits; equivalently, we could remove particles from the DF using a more gradual cut-off in energy, rather than an abrupt truncation. Additionally, we only consider mass loss in discrete steps once per orbit, at the time of pericentric passage. Continuous models for tidal mass loss are often created  by dividing the orbital period into discrete steps and assuming a fraction of the mass outside the tidal radius is lost at each of these steps, according to a characteristic time scale for mass loss \citep[e.g.][]{taylor2001a}. 
We found in Paper~II that a continuous mass-loss model of this kind requires an additional free parameter, calibrated to simulations, and thus we decided to use the discrete orbit-averaged model instead. However, this is clearly a simplification, and we aim to improve the mass-loss model in future work.

While our findings support the assertion that our energy-truncation model is universal, we have only considered spherical, isotropic subhaloes, evolving in an NFW main potential. Examining more complicated ICs (including anisotropic and multicomponent systems) will be the focus of future work. 
Another interesting extension would be to include a disk potential to the host halo, as it has been shown that including an embedded central disk potential to dark-matter-only simulations increases the rate of subhalo disruption \cite[e.g.][]{garrison2017}. 
For example, recent work by \cite{webb2020} considers more realistic external tidal fields, including baryonic bulge and disk components, and finds that the baryonic components lead to more subhalo disruption and lower mass subhaloes. Since the energy-truncation model works well for a wide range of collisionless systems, it may also be useful for predicting the evolution of multiple-component systems. For instance, tidal stripping is thought to be a likely mechanism for creating observed dark-matter-deficient ultra-diffuse galaxies \citep[e.g.][]{ogiya2021,ogiya2021b}. In future work, we will examine the evolution of these galaxies in energy space. 

%Summary
Overall, a growing body of evidence suggests that the central density and concentration of dark matter haloes are not well understood. Physically-based models for subhalo evolution are critical for correctly predicting density profile evolution at small radii, as these regions cannot be resolved in simulations. 
The theoretically-based energy-truncation model offers a convenient tool to study the evolution of collisionless systems as they are tidally stripped, allowing us to work towards a complete understanding of how subhaloes evolve, and to ultimately place more robust and accurate constraints on the properties of dark matter.

%%%%%%%%%%%%%%%%%%%%%%%%%%%%%%%%%%%%%%%%%%%%%%%%%%
%%%%%%%%%%%%%%%%%%%%%%%%%%%%%%%%%%%%%%%%%%%%%%%%%%
%%%%%%%%%%%%%%%%%%%%%%%%%%%%%%%%%%%%%%%%%%%%%%%%%%	

\section*{Acknowledgements}
The authors thank the anonymous referee for useful comments. NED acknowledges support from NSERC Canada, through a postdoctoral fellowship. JET acknowledges financial support from NSERC Canada, through a Discovery Grant. The simulations for this work were carried out on computing clusters provided by Compute Ontario (https://computeontario.ca/) and Compute Canada (www.computecanada.ca). 

\emph{Software}: \textsc{Gadget-2} \citep{gadget2}, \textsc{Icicle} \citep{drakos2017}, numpy \citep{numpy}, matplotlib \citep{matplotlib},  scipy \citep{scipy}.

\section*{Data availability}
	
The data underlying this article will be shared on reasonable request to the corresponding author.

\bibliographystyle{mnras}
\bibliography{TS3}

\appendix

%%%%%%%%%%%%%%%%%%%%%%%%%%%%%%%%%%%%%%%%%%%%%%%%%%	
\section{Accuracy of profile predictions}\label{sec:resid}
%%%%%%%%%%%%%%%%%%%%%%%%%%%%%%%%%%%%%%%%%%%%%%%%%%	

\begin{figure*}
	\centering
	\subfloat{{\includegraphics[trim={0 0cm 0 0cm},clip=true]{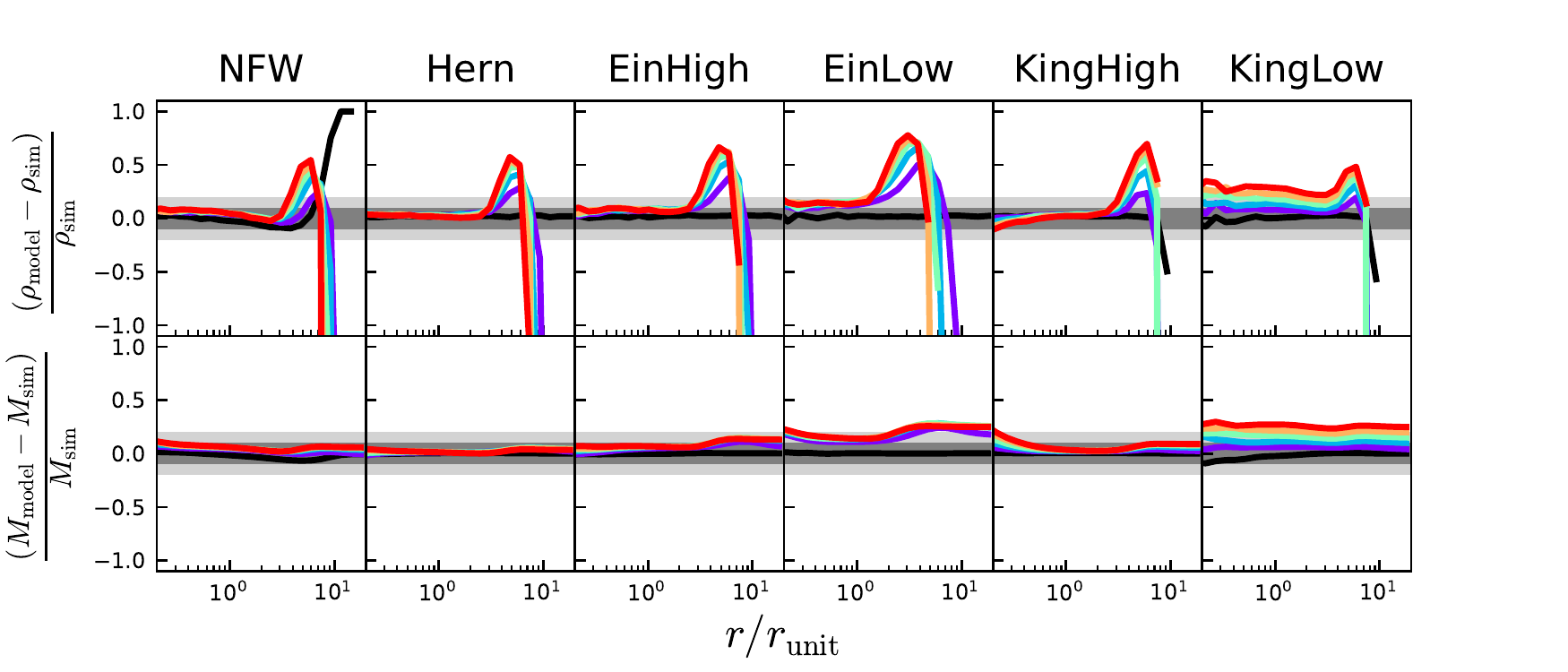}}}
	
	\subfloat{{\includegraphics[trim={0 0 0 0cm},clip=true]{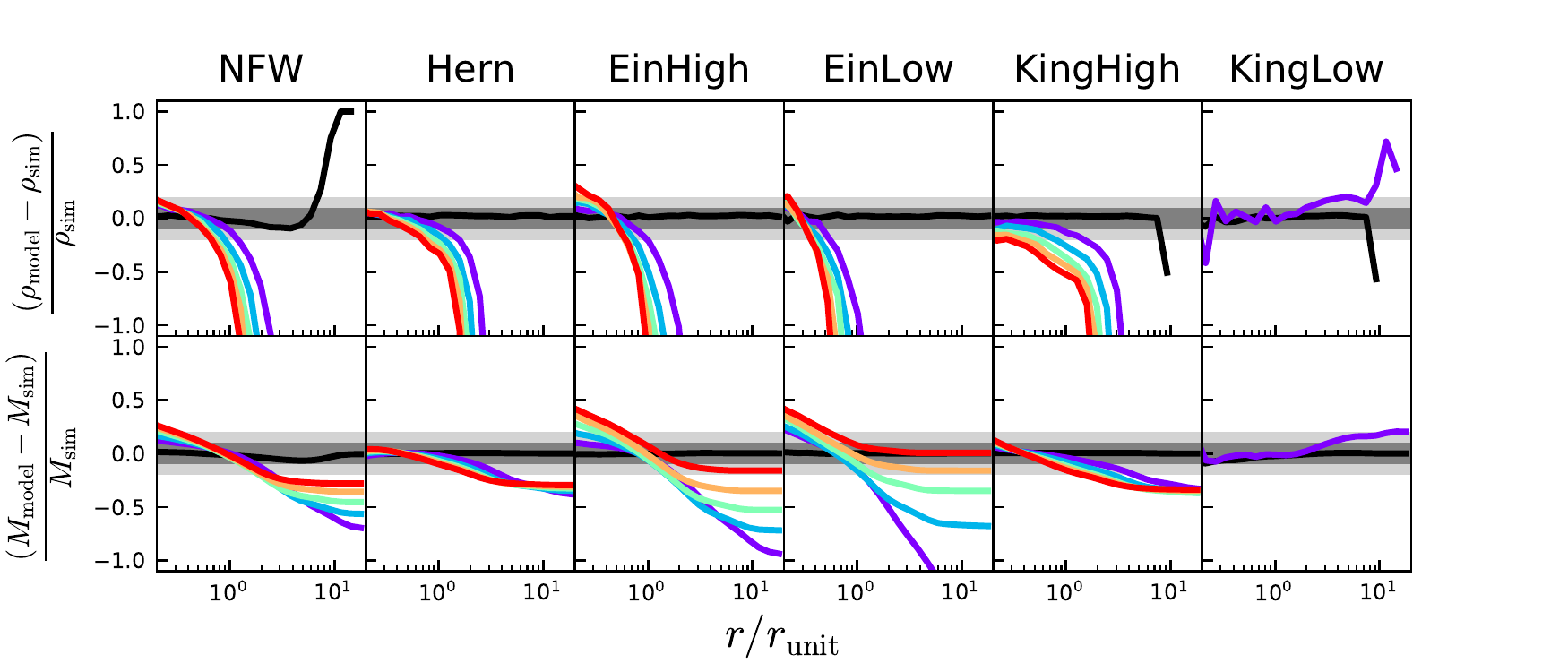}}}
	\caption{
	Residuals in the density profiles and mass profiles of the tidally stripped systems, when comparing the energy-truncation model to the Slow (top) and Fast (bottom) Simulations, as shown in Fig.~\ref{fig:ProfsFitsResid}. The shaded regions correspond to 10 per cent  (dark grey) and 20 per cent (light grey) agreement. In general, the model does a fairly good job at predicting the central density (within $\sim 2\, r_{\rm unit}$). At larger radii, there are significant deviations between the energy-truncation model and the self-bound remnant from simulations. The exact origin of these deviations are unclear, but are likely tied to the detailed mass-loss timescales of each particle, and how the bound particles of the satellite are defined.
	}
	\label{fig:ProfsFitsResid}
\end{figure*}

Fig.~\ref{fig:ProfsFitsResid} shows the residuals in the density and mass profiles of the energy-truncation model compared to the simulation results. Residuals are calculated as the difference between the model predictions and simulations, divided by the model predictions. We note that there is very little mass within $\sim r_{\rm unit}$---particularly for the EinLow and KingLow simulations---which makes the relative errors in the mass profiles untrustworthy at low radii. Nonetheless, these residual plots offer important insights into the accuracy of the energy-truncation model.

The Slow Simulation (top), generally captures the model to within $\sim 10$ per cent, except when $r$ approaches the truncation radius. An exception is the KingLow profile, whose central density is very sensitive to the total mass of the system. In general, for the Slow Simulation, the energy-truncation model over-predicts central density and mass of the stripped satellite. The Fast Simulations (bottom), shows a larger disagreement with the energy-truncation model (but the central density is still fairly well predicted, to within $\sim 20$ per cent or better). Since this simulation was chosen as a case where the energy-truncation model does not work as well (see Papers I and II), this is expected. In general, large deviations between the model and simulations can be seen at radii larger than $\sim 2\, r_{\rm unit}$. At these larger radii, the energy-truncation model predicts too much mass for the Slow Simulation, and too little mass for the Fast Simulation. This result is consistent with the mass-loss curves shown in Fig.~\ref{fig:MassLossCurves}. 

This paper has mainly focussed on predicting the central density of satellites, which the energy-truncation model does fairly well. The material at large subhalo radii are more dynamically complicated, and are sensitive to how we define bound particles. Future work will focus on understanding the detailed time scales of mass loss for all particles, and whether some of the bound mass is in the process of leaving the system.

%%%%%%%%%%%%%%%%%%%%%%%%%%%%%%%%%%%%%%%%%%%%%%%%%%	
\section{The deforming bowl picture of tidal mass loss}\label{sec:boost}
%%%%%%%%%%%%%%%%%%%%%%%%%%%%%%%%%%%%%%%%%%%%%%%%%%	

\cite{stucker2021} recently proposed that there is a natural energy at which particles will be lost, calculated from the ``boosted potential". They advocate for a model for tidal stripping in which mass loss is explained by a lowering of the escape energy caused by the tidal field termed the ``deforming bowl” picture. This naturally leads to the concept of a ``truncation energy".

\begin{figure}
	\includegraphics[width = \columnwidth]{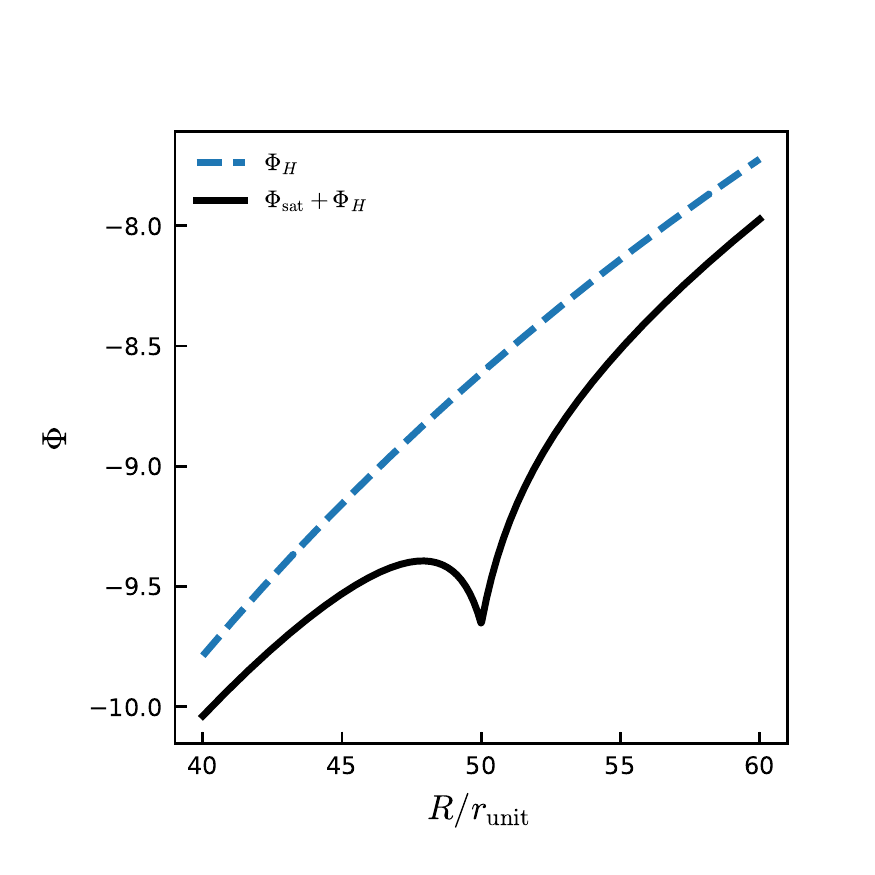}
	\caption{
	The global potential around an NFW satellite located at $R=50\, r_{\rm unit}$ of its host halo. The potential from the host halo strongly effects the local potential of the satellite.
	}
	\label{fig:GlobalPotential}
\end{figure}

In tidal stripping analyses, we often only consider the self potential of the satellite. However, the satellite exists within the large scale potential of the host it is orbiting,  as illustrated in Fig.~\ref{fig:GlobalPotential}. This figure shows an (infinitely extended) NFW potential of mass $M_{\rm sat}$ located at $R_P=50\,r_{\rm unit}$ of the host halo ($M_H = 300 M_{\rm sat}$).

To account for the affect of the host potential on the satellite, \cite{stucker2021} define the boosted potential as:
\begin{equation}
\Phi_{\rm boost} (\mathbf{x}) = \Phi (\mathbf{x}) + \mathbf{x} \cdot \mathbf{a_0}
 \end{equation}
where $\Phi$ is the total (host plus satellite) potential and $\mathbf{a_0}$ is an additional apparent acceleration. Thus, in our example, the boosted potential subtracts the local large scale gradient of the host halo from the total potential. 

Under the assumption of spherical symmetry for both the host and satellite halo, and assuming that the satellite is located at pericentre, \cite{stucker2021} define the boosted potential of the satellite system as:
\begin{equation} \label{eq:boost}
\begin{aligned}
\Phi_{\rm boost} (R) = &\Phi_H(R)+ \Phi_{\rm sat}(r)\\ & - (R-R_p) \dfrac{GM_H(<R_p)}{R_p^2} - \Phi_H(R_p)  \,\,\, ,
\end{aligned}
\end{equation}
where $R= R_p + r$. The final term has been subtracted to give the self-potential of the satellite. Solving for the saddle points of Equation~\eqref{eq:boost} gives:
\begin{equation} \label{eq:rhosat_boost}
\bar{\rho}_{\rm sat} (r) = \pm  \left[ \dfrac{R_p}{r} -\dfrac{R_p^3}{r (R_p \pm r)^2} \dfrac{M_H(< R_p \pm r)}{M_H(<R_p)} \right] \bar{\rho}_H(R_p) \,\,\, .
\end{equation}

\begin{figure}
	\includegraphics[width = \columnwidth]{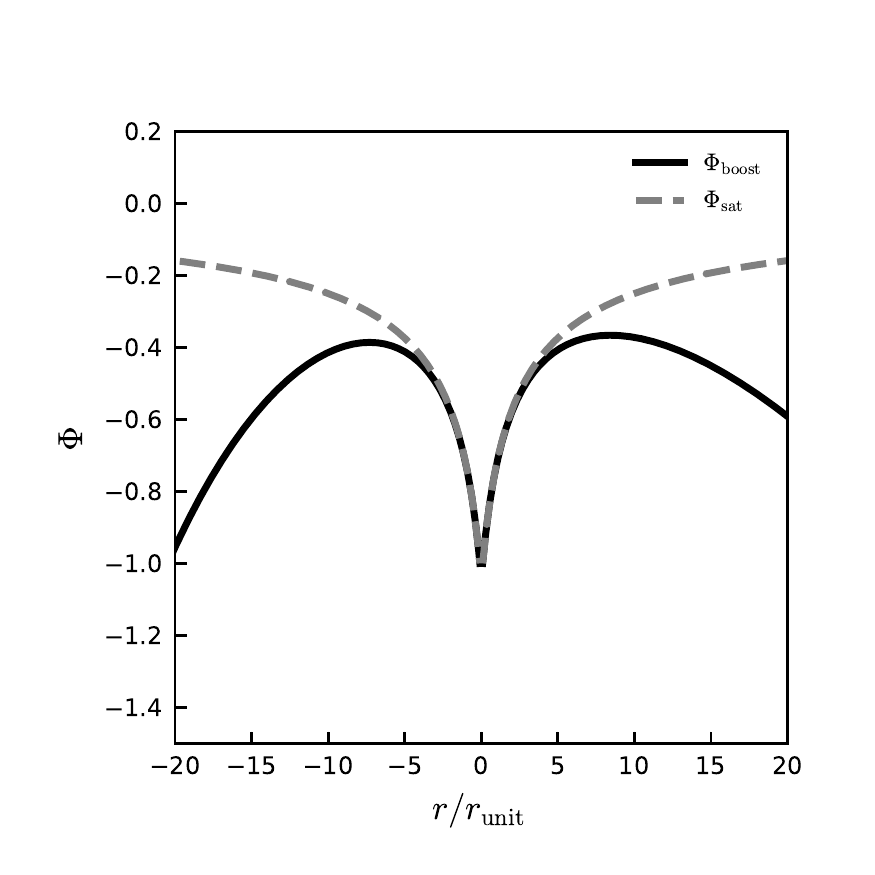}
	\caption{The potential of an NFW satellite (dashed grey line) and the boosted potential (solid black line). In the ``deforming bowl" picture, particles outside the saddle points of the boosted potential are lost.
	}
	\label{fig:BoostedPotential}
\end{figure}

Fig.~\ref{fig:BoostedPotential} shows the boosted potential of an NFW satellite. For this example, the saddle points are located at $r=-7.3\, r_{\rm unit}$ and  $8.4\, r_{\rm unit}$. In the deforming bowl picture, particles are  lost as they spill over the top of this bowl. 

This picture of mass-loss can be easily integrated into our energy-truncation model with the observation that Equation~\eqref{eq:rhosat_boost} is in the form of Equation~\eqref{eq:rlim}, with 
\begin{equation} \label{eq:eta_boost}
\eta_{\rm boost} = \pm \left( \dfrac{R_p}{r_{\rm lim}} - \dfrac{R_p^3}{r (R_p \pm r_{\rm lim})^2} \dfrac{M_H(< R_p \pm r_{\rm lim})}{M_H(<R_p)} \right) \,\,\,.
\end{equation}

Other common definitions of $\eta$, as discussed in detail in Paper II, include the well-known Roche ($\eta=2$) and Jacobi limits ($\eta= 3$), 
\begin{equation} \label{eq:eta1}
\eta_1 = 2 - \dfrac{ {\rm d} \ln M}{{\rm d} \ln R} \,\,\, ,
\end{equation}
which describes a satellite and host with extended bodies, and
\begin{equation}\label{eq:eta2}
\eta_2 = \dfrac{\omega^2}{\omega_c^2} - \dfrac{1}{\omega_c^2} \dfrac{{\rm d}^2 \Phi}{{\rm d} R^2} \,\,\, ,
\end{equation}
which includes the centrifugal force.  

In Paper II we found that in the energy-truncation framework, $\eta_2$ overestimates mass-loss rates and work the best for circular orbits, while  $\eta_1$ underestimates mass loss and work the best for radial orbits. The value we adopt, $\eta_{\rm eff}$, is the orbital average of the instantaneous value of $\eta_2$ (Equation~\eqref{eq:etaeff}). In practice, we found $\eta_{\rm eff}$ (which lies between $\eta_1$ and $\eta_2$) worked best to describe mass-loss.

There are important differences between Equation~\eqref{eq:eta_boost} and the alternate $\eta$ definitions listed above. First, there are two $\eta_{\rm boost}$ values, reflecting the fact that the tidal field is stronger on the side of the satellite closer to the host halo. We will take $\eta_{\rm boost}$ to be the negative root, as this corresponds to  stronger tidal forces. Secondly, since the value of $r_{\rm lim}$ depends on the satellite model, $\eta_{\rm boost}$ does not only depend on the host halo and orbit, but also on the details of the satellite profile. 
 
\begin{figure}
	\includegraphics[width=\columnwidth]{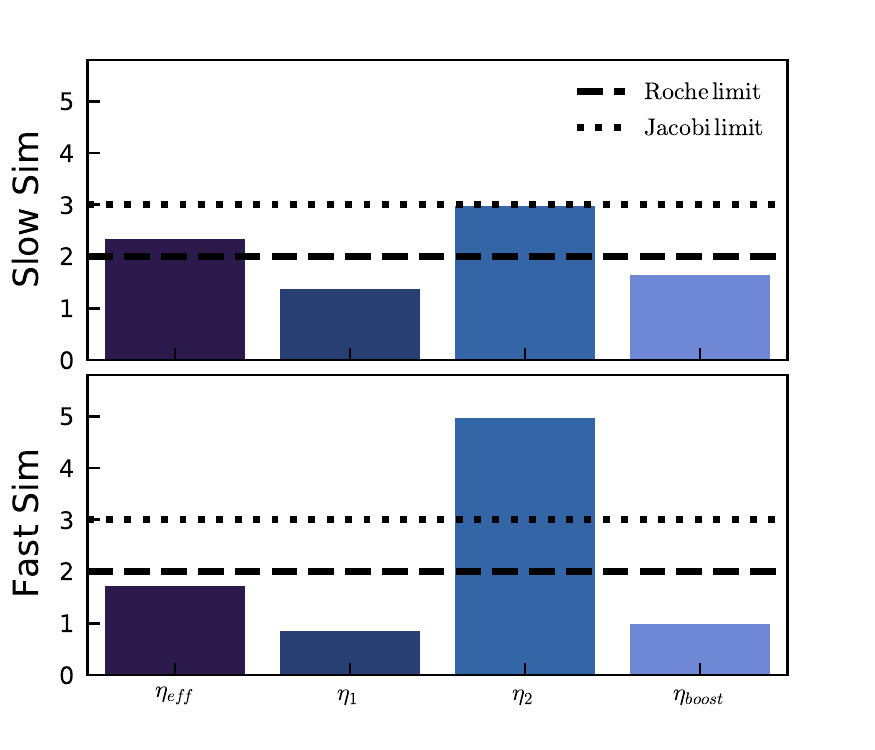}
	\caption{
Values of $\eta$ used in the mass-loss prescription, Equation~\eqref{eq:Mbnd}. From left to right (1) $\eta_{\rm eff}$ is the value we use in Equation~\eqref{eq:etaeff} shown to match well with simulations (Paper II) so can be considered close to the "true" value (2) $\eta_1$, Equation~\eqref{eq:eta1}, which works well for radial orbits (3) $\eta_2$,  Equation~\eqref{eq:eta2}, which works well for circular orbits and (4) $\eta_{\rm boost}$, Equation~\eqref{eq:eta_boost}, which is calculated from the boosted potential. For comparison, we also show the Roche ($\eta=2$) and Jacobi ($\eta=3$) limits.  $\eta_{\rm boost}$ is lower than the ``true" value, $\eta_{\rm eff}$, and therefore under-predicts mass loss in the energy-truncation framework.}
	\label{fig:EtaCompare}
\end{figure}

In Fig.~\ref{fig:EtaCompare} we compare $\eta_{\rm boost}$ to $\eta_{\rm eff}$ as well as commonly used values in the literature. We consider both the Slow and Fast Simulation, and use an infinitely extended NFW profile to calculate the $\eta_{\rm boost}$ values. We find that the boosted potential calculation predicts an $\eta$ value between   $\eta_1$ and $\eta_{\rm eff}$. Since we interpret $\eta_{\rm eff}$ as being the ``true" value, this suggests that $\eta_{\rm boost}$  underestimates mass-loss in our framework. 

Although $\eta_{\rm boost}$ (and effectively the truncation energy) calculated from this formulation of the boosted potential underestimates the best-fit value (which is expected to be roughly equivalent to $\eta_{\rm eff}$), the boosted potential framework seems to be a natural extension to the energy-based tidal stripping model we advocate. A potential explanation for why $\eta_{\rm boost}$ is too low is because we neglected the centrifugal term; i.e. we could subtract a term  $\frac{1}{2} |\omega \times \mathbf{R}|^2$ from Equation~\eqref{eq:boost} to account for the corotating potential. We leave more detailed explorations of the boosted potential to future work.

%%%%%%%%%%%%%%%%%%%%%%%%%%%%%%%%%%%%%%%%%%%%%%%%%%
%%%%%%%%%%%%%%%%%%%%%%%%%%%%%%%%%%%%%%%%%%%%%%%%%%
%%%%%%%%%%%%%%%%%%%%%%%%%%%%%%%%%%%%%%%%%%%%%%%%%%	
	% Don't change these lines
	\bsp	% typesetting comment
	\label{lastpage}
\end{document}